\documentclass[aps,pra,reprint,superscriptaddress,floatfix]{revtex4-1}
\usepackage{float}
\usepackage[T1]{fontenc}
\usepackage[utf8]{inputenc}
\usepackage[british,UKenglish,USenglish,english,american]{babel}
\usepackage{graphicx}
\usepackage{rotating}
\usepackage{amsmath}
\usepackage{color}
\usepackage{bm}
\usepackage{booktabs}
\renewcommand{\BibitemShut}[1]{}

\begin{document}

\author{Juha Vaara}
\affiliation{NMR Research Unit, P.O.~Box 3000, FI-90014 University of Oulu, Finland}
\email{juha.vaara@iki.fi} 
\author{Michael~V.~Romalis}
\affiliation{Department of Physics, Princeton University, Princeton, New Jersey, 08544, USA}

\title{Precision Calculation of Scalar Nuclear Spin-Spin Coupling in a Noble Gas Mixture}

%\begin{tocentry}
%\begin{center}
%\includegraphics[width=6.5cm]{toc.jpg}
%\end{center}
%\end{tocentry}

\begin{abstract}
Indirect  spin-spin interactions between nuclei bound in a molecule are well-known in nuclear magnetic resonance, but such interactions between unbound atoms are less well studied and are often assumed to be zero.  We present the first precision calculation of this interaction between $^{129}$Xe and $^3$He nuclei in a gas. Relativistic,
state-of-the-art electronic structure theory is used to compute the scalar coupling constant $J\!\left(R\right)$  and the interatomic potential energy function, $V\!\left(R\right)$, as functions of the Xe-He internuclear distance $R$. Using virial expansion we find the nuclear spin enhancement factor $\kappa= -0.0105 \pm 0.0015$ in excellent agreement with recent experiments. This interaction is particularly important for precision measurements using nuclear spin co-magnetometers.
\end{abstract}

\keywords{noble gas, co-magnetometer, nuclear magnetic resonance, spin-spin coupling, electronic structure}

\maketitle

%\maketitle

%\section{Introduction}
%\noindent T\"ast\"a l\"ahtee

\section{Introduction}

Nuclear spins interact directly via magnetic moments and indirectly via electrons. In a liquid or gas the direct magnetic dipolar and the anisotropic part of indirect interactions are averaged out, leaving only the scalar spin-spin coupling, SSC ($J$ coupling). SSC serves as one of the main analysis tools of nuclear magnetic resonance (NMR) spectroscopy and is routinely measured and 
%theoretically 
calculated in molecules~\cite{lev01,helgaker_ab_1999}. In contrast, SSC between unbound nuclear spins, {\em e.g.}, in van der Waals (vdW) complexes, have been analyzed theoretically for only a few systems~\cite{sal98,pec00,pec01,bag01,bag02,vaa13}. In such cases SSC is also much harder to measure because it is normally averaged by molecular motion. SSC is expected to be significantly smaller than the direct  interaction. There has been only one experimental observation of SSC for unbound spins, using a liquid mixture of Xe and pentane~\cite{led12}. The complex nature of this 
%van der Waals 
system 
allowed only a qualitative comparison with theory. In one case, the influence of the anisotropy of SSC could be inferred in a system with Xe atoms confined to a zeolite cage~\cite{jok13}.

In this work we present the first precision calculation of SSC in a vdW complex, of $^{129}$Xe and $^3$He, which can be compared with recent 
experimental observations~\cite{lim18,ter18}.
%measurements~\cite{lim18,ter18}. 
The calculation can done with a precision comparable to or better than for the much larger interaction between unpaired electron and nucleus, as has been performed for alkali metal-noble gas  pairs, {\em e.g.}, in Refs.~\onlinecite{walker_1989,han17}. We find that SSC between nuclear spins is a general feature. For example, we estimate that for the $^{129}$Xe-$^{131}$Xe pair, the nuclear spin
enhancement 
factor $\kappa$ of the associated frequency shift~\cite{walker_1997} is $-0.35$, almost approaching the strength of the direct magnetic dipolar 
coupling, which in a simple classical picture would correspond to $\kappa=1$.

While SSC between unbound spins is small, typically resulting in frequency shifts of less than 1~Hz, it has significant impact on precision measurements using nuclear spin co-magnetometers that seek to measure nHz-level shifts due to new particle physics phenomena~\cite{saf18}.  In particular, noble gas systems can be used to search for dark matter particles~\cite{gra18},
long-range spin-dependent forces~\cite{gle08,bul13,tul13}, as well as tests of CP-~\cite{ros01}, Lorentz- and CPT-symmetries~\cite{bea00}.
As the signals involved are small, it is essential
to reliably estimate and eliminate frequency
shifts and drifts due to conventional phenomenology and
instrumentation, including SSC~\cite{hec03}.
For SSC 
%quantum-chemical 
calculations to be useful in 
%the 
precision magnetometry, 
%community, 
it is essential to account well for both electron correlation and, due to heavy elements, relativistic effects~\cite{rei09}.
% on electronic structure~\cite{rei09}. 
In particular, in four-component relativistic 
%description of the electronic structure 
theory, one single magnetic hyperfine operator~\cite{mos73} replaces the four interactions of the nonrelativistic (NR) approach: the Fermi contact and dipolar Hamiltonians coupling nuclear and electron spins, as well as two orbital hyperfine Hamiltonians coupling the nuclear spin(s) to electron motion. 
%Indeed, in four-component relativistic 
%description of the electronic structure 
%theory, only one hyperfine operator prevails~\cite{mos73}, and t
The isotropic, rank-0 tensorial component of the resulting total relativistic SSC
%, instead of the Fermi contact part as in nonrelativistic (NR) theory,
determines 
%the modification
%enhancement 
the  $\kappa$ factor of the associated frequency shift.
% in the gas phase~\cite{walker_1997}.
 
Here, relativistic first-principles electronic structure theory is used to estimate the frequency shift occurring in a $^{129}$Xe-$^3$He co-magnetometer as a result of the indirect, electron-mediated SSC between the two isotopes. The treatment is based on the leading interaction term in the semiclassical virial expansion of the SSC constant in low-pressure gas mixture~\cite{buckingham_1956}. The second virial coefficient, $J_{1}$, appearing in this term is determined by SSC and potential energy (PEC) curves, $J\!\left(R\right)$ and $V\!\left(R\right)$, respectively, as functions of the interatomic distance $R$. These curves are 
%presently 
calculated using state-of-the-art quantum-chemical tools. The enhancement factor $\kappa$ of $^{129}$Xe frequency, as well as its temperature dependence are obtained from 
%the second virial coefficient 
$J_{1}$. The result, $\kappa = -0.0105$ agrees very well with experiments~\cite{lim18,ter18}. There is also an order-of-magnitude agreement with earlier experimental $\kappa_{\rm XeH} = -0.0014$~\cite{led12} for $^1$H-$^{129}$Xe SSC over vdW bonds between 
%atomic 
Xe and the protons of pentane.

\section{Theory}

SSC constant experienced by 
%the 
$^{129}$Xe 
%isotope 
with the surrounding $^3$He nuclei in the gas mixture can be obtained from the virial expansion at temperature $T$~\cite{buckingham_1956}
\begin{equation}
J_{\rm XeHe}\!\left(n_{\rm He},T\right) = J_{1}\!\left(T\right)n_{\rm He} + J_{2}\!\left(T\right)n_{\rm He}^2\ldots,
\label{eqn-virial}
\end{equation}
where $n_{\rm He}$ is the number density of 
%the 
$^3$He.
%gas 
%component 
%and $T$ is temperature. 
In low-pressure gas, SSC 
%constant 
can be approximated from the first term, in which the second virial coefficient 
%(the first coefficient,~
($J_{0}$ vanishes for a pair interaction quantity such as $J$) 
%of SSC constant between $^{129}$Xe and $^3$He nuclei, 
is obtained as
\begin{equation}
J_{1}\!\left(T\right)=4\pi\int_{0}^\infty J\!\left(R\right)\,\exp\left[-V\!\left(R\right)/kT\right] R^2\,dR.
\label{eqn-j1}
\end{equation}
The Zeeman interaction of the $^{129}$Xe nucleus is described by the spin Hamiltonian
%\begin{equation}
$H_{\rm Z} = -\gamma_{\rm Xe}\hbar \bm{I}_{\rm Xe}\cdot
%\left(\bm{1}-\bm{\sigma}_{\rm Xe}\right)\cdot
\bm{B}_{\rm Xe}$,
%\end{equation}
where $\bm{\mu}_{\rm Xe} = \gamma_{\rm Xe}\hbar \bm{I}_{\rm Xe}$ is the magnetic moment of $^{129}$Xe, expressed in terms of the gyromagnetic ratio $\gamma_{\rm Xe}$ and spin vector $\bm{I}_{\rm Xe}$ of the isotope. 
%$\bm{B}_{0}$ is the externally applied, homogeneous magnetic field and 
$\bm{B}_{\rm Xe}$ 
%= \left(\bm{1}-\bm{\sigma}_{\rm Xe}\right)\cdot\bm{B}_{0}$ 
is the effective field at 
%the site of 
the $^{129}$Xe nucleus.
%, parameterized by the nuclear shielding tensor $\bm{\sigma}_{\rm Xe}$. 
The Hamiltonian of the 
%indirect 
$^{129}$Xe-$^3$He SSC is, in turn,
%\begin{equation}
$H_{J} = h \bm{I}_{\rm Xe}\cdot\bm{J}_{\rm XeHe}\cdot\bm{I}_{\rm He}$,
%\end{equation}
with the SSC tensor $\bm{J}_{\rm XeHe}$ and $^3$He spin, $\bm{I}_{\rm He}$. The rank-0 part of $\bm{J}_{\rm XeHe}$ causes an extra contribution $dB_{\rm Xe}$ to the effective field:
\begin{eqnarray}
-\gamma_{\rm Xe}\hbar I_{\rm Xe}dB_{\rm Xe}&=&hI_{\rm Xe}J_{\rm XeHe}\langle I_{\rm He}\rangle \nonumber \\
\Rightarrow dB_{\rm Xe}&=&-\frac{2\pi}{\gamma_{\rm Xe}}J_{\rm XeHe}\langle I_{\rm He}\rangle,
\label{eqn-db-mic}
\end{eqnarray}
in an isotropic sample, with the average $^3$He spin $\langle I_{\rm He}\rangle$.

The corresponding field increment can be formulated as arising from the assumed uniform magnetization $M_{\rm He}$ of a spherical sample of $^3$He nuclei around $^{129}$Xe~\cite{schaefer_1989}
%the xenon atom~\cite{schaefer_1989}
\begin{equation}
dB_{\rm Xe}=\kappa\frac{2\mu_{0}}{3}M_{\rm He}=\kappa\frac{2\mu_{0}}{3}n_{\rm He}\gamma_{\rm He}\hbar\langle I_{\rm He}\rangle,
\label{eqn-db-emp}
\end{equation}
where $\kappa$ represents the modification due to the electron cloud of Xe. 
%Equating the two expressions for $dB_{\rm Xe}$, 
From Eqs.~(\ref{eqn-db-mic}--\ref{eqn-db-emp}), 
%provides the expression
\begin{equation}
\kappa=-\frac{3\pi}{\mu_{0}\hbar}\frac{1}{\gamma_{\rm He}\gamma_{\rm Xe}}\frac{J_{\rm XeHe}}{n_{\rm He}} =-\frac{3\pi}{\mu_{0}\hbar}\frac{1}{\gamma_{\rm He}\gamma_{\rm Xe}}J_1,
\label{eqn-kappa}
\end{equation}
where the first term of the expansion of Eq.~(\ref{eqn-virial}) has been used. $\kappa$ is %seen to be 
determined by the second virial coefficient of $^{129}$Xe-$^3$He SSC constant, $J_{1}$.

\section{Calculations of potential curve}

First-principles calculations of $J_{1}$ as defined by Eq.~(\ref{eqn-j1}) were pursued using a combination of electronic structure tools. $V\!\left(R\right)$ was calculated at 54 internuclear distances ($R=2.4$~\AA\ldots 7.3~\AA\ in steps of 0.1~\AA, and $R=7.3$~\AA\ldots 8.1~\AA\ in steps of 0.2~\AA) at the coupled-cluster singles, doubles and perturbative triples [CCSD(T)] level using the {\sc Molpro} code~\cite{wer12,molpro}. The calculations used the Gaussian aug-cc-pV6Z basis set for He~\cite{dun89}, as well as the energy-consistent relativistic ECP28MDF pseudopotential (with 28 electrons 
%retained 
in the atomic core) and the corresponding aug-cc-pV5Z-PP valence basis 
%set 
for Xe~\cite{peterson_2003}. To ensure basis-set saturation in the interatomic spatial region, two measures were taken. First, 
%the 
Xe 
%center 
was furnished with three 
%extra primitive, 
diffuse 
%(small-exponent) 
Gaussian exponents in each of the $spdfgh$-shells. These exponents 
%of these functions 
were derived by successive division by a factor of three from the most diffuse function in the aug-cc-pV5Z-PP basis in each angular momentum shell. Secondly, further primitive functions were added at the mid-point of the Xe--He ``bond''. The bond functions (BFs) had exponents 0.9, 0.3, and 0.1 for the three $s$- and $p$-type primitives, 0.6 and 0.2 for the two $d$- and $f$-type primitives, and 0.35 for one $g$-type primitive~\cite{slavicek_2003}. The counterpoise correction~\cite{boys_1970} was applied to reduce the basis-set superposition error.

\begin{figure}[tb]
\includegraphics[width=0.9\columnwidth]{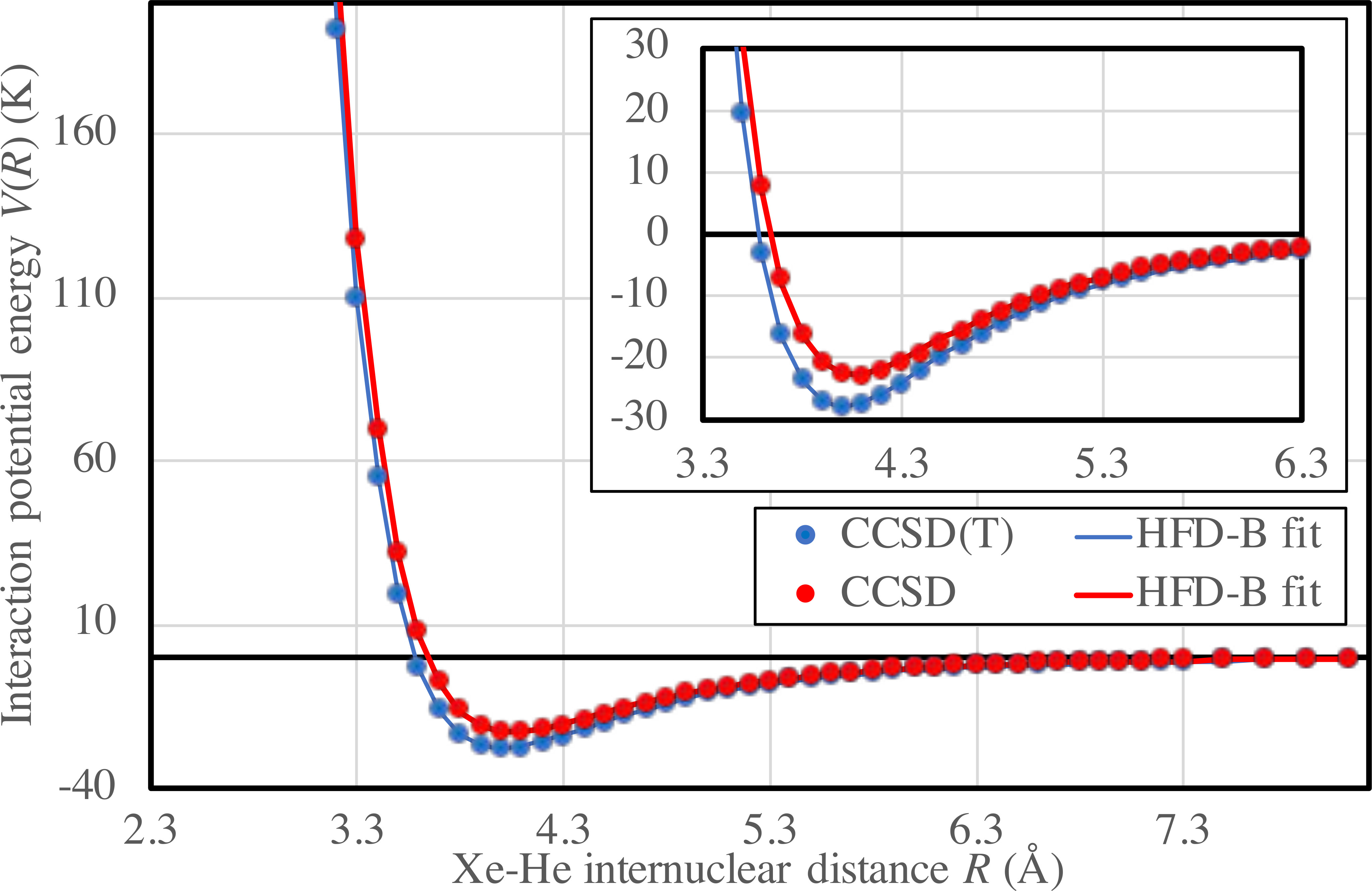}
\caption{Calculated CCSD(T) potential energy $V\!\left(R\right)$ for the Xe--He dimer as a function of the internuclear distance $R$. Fit to the HFD-B form of PEC 
%[Eq.~(\ref{eqn-v})] 
is also shown. For comparison, $V\!\left(R\right)$
%the potential energy function 
based on the simpler CCSD level of theory is included. Inset shows a close-up of the well region.}
\label{fig-v}
\end{figure}

Fig.~\ref{fig-v} shows the calculated 
%CCSD(T)-level 
$V\!\left(R\right)$ along with its fit to the analytical Hartree-Fock-dispersion (HFD-B) 
%energy 
form~\cite{azi77}.
%\begin{eqnarray}
%V\!\left(R\right) & = & \epsilon V^\star\!\left(x\right) \hspace{0.5cm};\hspace{0.5cm}x=R/R_{m}\nonumber \\
%V^\star\!\left(x\right) & = & L\exp\left(-\gamma x+\beta x^2\right)-F\!\left(x\right)\sum_{j=0}^2c_{2j+6}/x^{2j+6} \nonumber \\
%F\!\left(x\right)&=& \left\{\begin{tabular}{ll}
%$\exp\left[-\left(\frac{D}{x}-1\right)^2\right]$, & $x<D$ \\
%$1$, & $x\geq D$ \\
%\end{tabular} \right.
%\label{eqn-v}
%\end{eqnarray}
The formula and the fit parameters are given in the Supplemental Material~\cite{supp}.
%\begin{table}[h!]
%\caption{Fit parameters of the CCSD(T) and CCSD potential energy functions for the Xe--He system.
%% into the form of Eq.~(\ref{eqn-v}).
%}
%\label{tbl-v}
%%\vspace{0.5cm}
%\begin{tabular}{lrr}\hline\hline
%Parameter & CCSD(T) & CCSD \\ \hline
%$\epsilon$ (K) & 24.526136 & 24.871021 \\
%$R_{m}$ (\AA)  &  4.175781 & 4.087755 \\
%$L$            & 34253.231137 & 34459.321089 \\
%$\gamma$       & 5.008183 & 4.906990 \\
%$\beta$        & $-6.572879$ & $-6.309467$ \\
%$c_{6}$        & 1.668871 & 1.612432 \\
%$c_{8}$        & $-0.477006$ & $-0.479358$ \\
%$c_{10}$       & $-0.044923$ & $-0.044957$ \\
%$D$            & $1.436190$ & $1.483476$ \\
%\hline\hline\end{tabular}
%\end{table}
The fitted PEC well depth is 28.0~K at 4.00~\AA. The relevant temperatures for 
%the SEOP processes and 
co-magnetometer operation start from roughly 300~K upwards, {\em i.e.}, with thermal energies significantly larger than the well depth, meaning that the semiclassical treatment of the virial coefficient in Eq.~(\ref{eqn-j1}) is well valid~\cite{riz02}. The calculated well depth and minimum-energy separation are in very good agreement with the results of the analysis of high-resolution crossed molecular-beam data in Ref.~\onlinecite{azi89}, $28.95\pm 0.46$~K and 3.975~\AA, respectively.
%, in the best-fit data reported in that paper. 
Fig.~\ref{fig-v} 
%and Table~\ref{tbl-v} 
also includes lower-level CCSD results ({\em i.e.}, without perturbative triples) in which the potential minimum is shallower (well depth 22.7~K) and occurs at larger internuclear distance ($R=4.05$~\AA) than what is seen in the CCSD(T) data. The difference between CCSD(T) and CCSD PECs is indicative of the remaining systematic error in $V\!\left(R\right)$.

\section{Calculations of spin-spin coupling curve}

%$J\!\left(R\right)$
The SSC calculations were carried out 
%was calculated 
at the same internuclear separations as used for $V\!\left(R\right)$. For $J\!\left(R\right)$, nor other second-order magnetic properties, there are no {\em ab initio\/} 
%wavefunction theory 
methods available that would meaningfully combine post-Hartree-Fock electron correlation and relativistic effects, however. The available four-component density-functional theory (DFT) methods for $J\!\left(R\right)$ based on the 
%``fully relativistic`` 
Dirac-Fock Hamiltonian~\cite{DIRAC17,respectJ} suffer from significant dependence of the results on the chosen exchange-correlation functional; indeed $J$ can be considered as one of the most challenging properties for DFT~\cite{vaa07}. Instead, we followed a piecewise strategy similar 
%in spirit 
to that used for $J\!\left(R\right)$ in xenon dimer in Ref.~\onlinecite{vaa13}. In this approach, the isotropic 
%SSC constant 
$J$ 
%(equal to one third of the trace of the $\bm{J}$ tensor) 
at each $R$ is obtained as a sum 
\begin{eqnarray}
J&\approx&{\rm CCSD(NR)}\nonumber \\
&& \mbox{} + {\rm SOPPA(CCSD)(NR\ w/ BF)} - {\rm SOPPA(CCSD)(NR)} \nonumber \\
&& \mbox{} + {\rm PBE0(R\ w/ BF)} - {\rm PBE0(NR\ w/ BF)}
\label{eqn-j}
\end{eqnarray}
of the results of distinct first-principles calculations. On the first row of Eq.~(\ref{eqn-j}),  NR CCSD-level data is obtained using the {\sc Cfour} code~\cite{CFOUR} and the Gaussian ccJ-pV5Z basis~\cite{ben08} for He, and the same completeness-optimized~\cite{man06}, uncontracted 27$s\,$25$p\,$21$d\,$4$f$ Xe basis as was used for the hyperfine interactions of $^{129}$Xe in Refs.~\onlinecite{han17,roukala_2015}. All the four NR contributions~\cite{helgaker_ab_1999} to $J$ --- the para- and diamagnetic nuclear spin-electron orbit,
% (PSO and DSO) terms, 
the spin-dipole 
%term (SD) 
and the Fermi contact terms,
% (FC), 
of which the last one overwhelmingly dominates 
%the NR $J$ 
in this system --- were included.
% in the calculation. The 
CCSD 
%method 
was adopted as the highest manageable level of theory for the present case, as CCSD(T) does not offer a similar benefit for $J$ as it does for PEC~\cite{aue01,aue09} and the cost of even higher-level methods~\cite{aue01,aue09} is prohibitive.

On the second row of Eq.~(\ref{eqn-j}), an attempt is made to remedy the remaining basis-set deficiency of the CCSD calculation, by carrying out two calculations (for each $R$) at the second-order polarization propagator (SOPPA) level with CCSD amplitudes [SOPPA(CCSD)]~\cite{ene98}, on the {\sc Dalton} code~\cite{dalton,aid14}. A difference is taken between results obtained with (BF) and without the same set of additional BFs, as used for $V\!\left(R\right)$. Program limitations inhibited us from using BFs in the full CCSD calculations.

Finally, on the third line of Eq.~(\ref{eqn-j}), relativistic corrections are applied as the difference between the 4-component Dirac-Kohn-Sham (R) DFT results using the PBE0 hybrid functional~\cite{adamo_1998,adamo_1999} obtained on the {\sc Dirac} code~\cite{DIRAC17} and the corresponding NR PBE0 calculation
%, using the same 
%PBE0 
%functional, 
on {\sc Dalton}. 
%The 
PBE0 
%functional 
was chosen as it gave, at the NR level where such comparison can be made, the closest agreement with the {\em ab initio\/} CCSD data for the $J\!\left(R\right)$ curve, among the tested functionals (PBE~\cite{perdew_1996}, PBE0, BLYP~\cite{becke_1988,lee_1988,miehlich_1989}, B3LYP~\cite{lee_1988,vosko_1980,becke_1993,stephens_1994}, BHandHLYP~\cite{lee_1988,becke_1993}, results not shown). For the large component of the relativistic wave function, the ccJ-pV5Z/27$s\,$25$p\,$21$d\,$4$f$ basis (for He/Xe) was used, but with also the He basis uncontracted, as well as supplemented with BFs as discussed above. The corresponding, uncontracted small-component basis was generated using unrestricted kinetic balance~\cite{sun11}. The NR PBE0 calculations used the large-component basis of the R PBE0 computations. 

\begin{figure}[tb!]
\includegraphics[width=0.9\columnwidth]{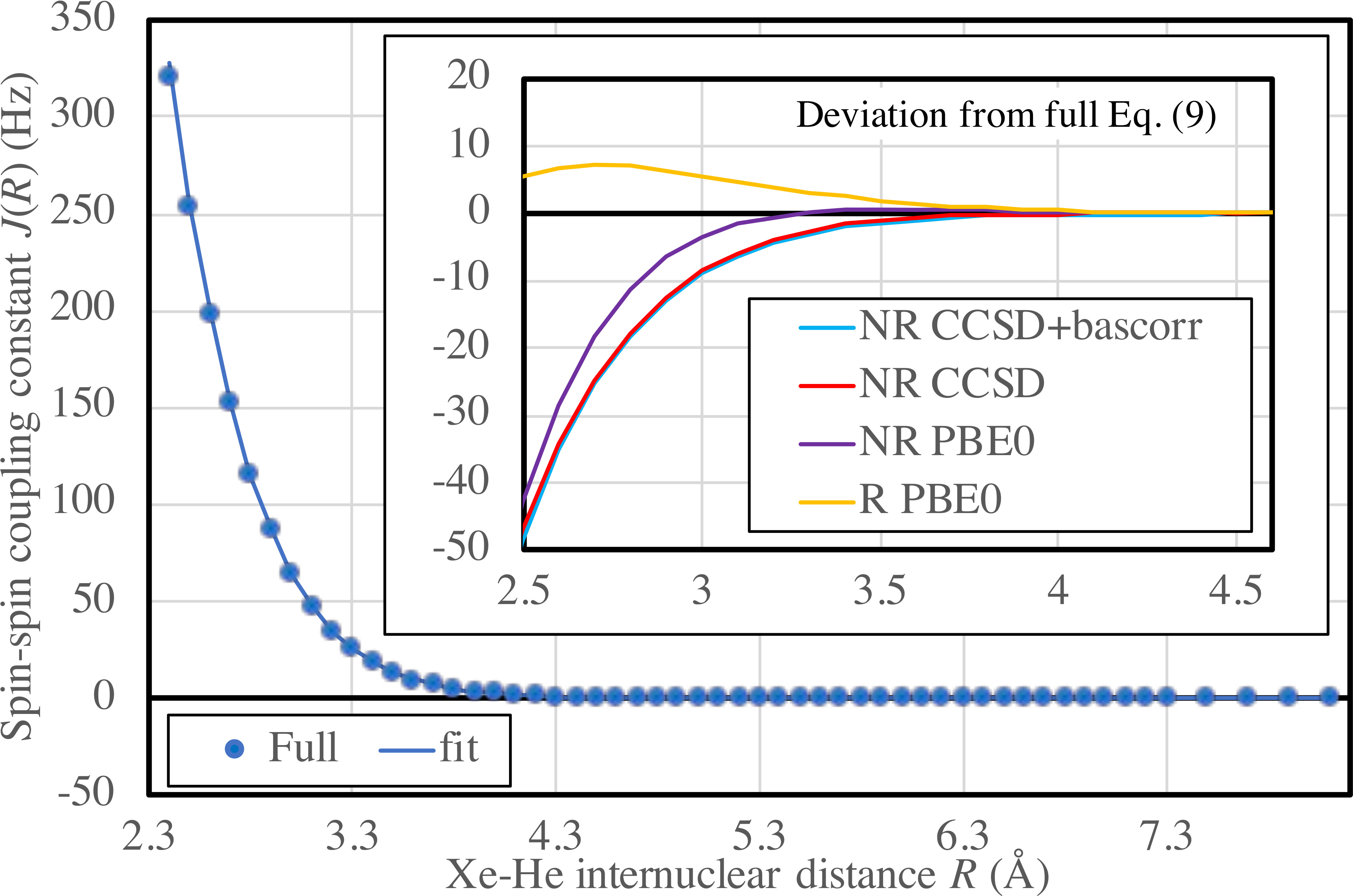}
\caption{Calculated $^{129}$Xe-$^3$He spin-spin coupling constant as a function of the internuclear separation $R$. The main panel shows the CCSD curve including both basis-set and relativistic corrections (Full) according to Eq.~(\ref{eqn-j}). The inset illustrates the difference from those data resulting from omitting the relativistic correction [last line of Eq.~(\ref{eqn-j}), ``NR CCSD+bascorr''], 
%both the relativistic correction 
and also the basis-set deficiency correction [second line of Eq.~(\ref{eqn-j}), ``NR CCSD'']. Corresponding deviations from the Full data resulting from  DFT/PBE0 calculations at the nonrelativistic (NR PBE0) and relativistic (R PBE0) levels, are also shown.}
\label{fig-j}
\end{figure}

Fig.~\ref{fig-j} shows the internuclear $J\!\left(R\right)$ curve as obtained using Eq.~(\ref{eqn-j}), as well as deviations thereof resulting from selected, more approximate levels of theory. The calculated data have been least-squares fitted to 
%the form
\begin{equation}
J\!\left(R\right)=\frac{C}{R^{p\left(R\right)}}\hspace{0.5cm};\hspace{0.5cm}p\!\left(R\right)=p_{0}+p_{1}R.
\label{eqn-fit}
\end{equation}
No physical interpretation of this form or the numerical parameters is implied---Eq.~(\ref{eqn-fit}) is merely used as a convenient form for interpolating the 
%quantum-mechanical 
data. To emphasize the physically relevant internuclear distance range, the magnitude of the function $w\!\left(R\right)=R^2J\!\left(R\right)\exp\left[-V\!\left(R\right)/kT\right]$ was used as the weighting factor, with the experimentally relevant temperature parameter $T=120$~C. The fit parameters are listed in the Supplemental Material~\cite{supp}.
%\begin{table}
%\caption{Fit parameters in Eq.~(\ref{eqn-fit}) 
%%corresponding to various computational approximations 
%of the interatomic $^{129}$Xe-$^3$He spin-spin coupling constant curve.}
%\label{tbl-j}
%%\vspace{0.5cm}
%\begin{tabular}{lrrr}\hline\hline
%& \multicolumn{3}{c}{Parameter} \\ \cline{2-4}		
%Level&$C$&$p_0$&$p_1$ \\ \hline
%Full Eq.~(9)&1902.188985&-2.277090&1.785515\\
%NR CCSD+bascorr&1005.102866&-2.865762&1.833998\\
%NR CCSD&1261.128732&-2.419153&1.751499\\
%NR PBE0&746.699666&-3.205819&1.828408\\
%R PBE0&1532.203488&-2.479622&1.762515\\
%\hline\hline\end{tabular}
%\end{table}

The SSC constant increases very steeply at small intermolecular distances, well below the minimum of the PEC. Frequency shifts due to interatomic SSC arise in deep-impact collisions of the interacting species. Both electron correlation and relativistic influences on  $J\!\left(R\right)$  are very significant in our best data, as specified by Eq.~(\ref{eqn-j}). The former aspect is seen from the comparison of the NR CCSD and PBE0 results, with DFT producing a larger SSC over the relevant range of internuclear distances $R$. The magnitude of the difference is about 4~Hz at $R=3.1$~\AA, which coincides with the maximum of the function $w\!\left(R\right)$. Relativistic contribution to SSC is positive and increases steeply towards smaller $R$. Its magnitude at $R=3.1$~\AA\ is over 6~Hz. 
%It is apparent, however, that t
The role of the basis-set correction obtained on the second line of Eq.~(\ref{eqn-j}) is very small, as the data show a very small difference between the ``NR CCSD'' and ``NR CCSD+bascorr'' results. This implies that already our original basis set (without the additional BFs) is quite well-converged for $J\!\left(R\right)$.

\section{Final results and discussion}

Table~\ref{tbl-kappa} reports the the calculated $J_{1}\!\left(T\right)$ according to 
%the semiclassical formula 
Eq.~(\ref{eqn-j1}), as well as $\kappa$ resulting from Eq.~(\ref{eqn-kappa}). Here, the CCSD(T) PEC 
%$V\!\left(R\right)$ 
%(with parameters given in Table~\ref{tbl-v}) 
and the piecewise approximated $J_{1}\!\left(R\right)$ from Eq.~(\ref{eqn-j}) were used, as best results available to us. In addition, the effects of systematic error in the computations of $V\!\left(R\right)$ and $J\!\left(R\right)$ were very conservatively estimated by repeating the calculations with the CCSD PEC and NR CCSD function for SSC, respectively. Error margins adopted from the resulting changes of $J_{1}$ and $\kappa$
% resulting from these, lower-level choices 
are also given in Table~\ref{tbl-kappa}. Our value for $\kappa$ equals $-0.0105\pm 0.0015$, with one third of the error  arising from 
%the choice of 
$V\!\left(R\right)$ and two thirds from 
%that of 
$J\!\left(R\right)$. There is excellent agreement with both the experiments of Refs.~\onlinecite{lim18,ter18};
% and \cite{ter18}; 
indeed the theory and experiments agree with each other to within the reported error margins.
\begin{table}[tb]
\caption{Second virial coefficient $J_{1}$ of the $^{129}$Xe-$^3$He spin-spin coupling and the enhancement factor of $^{129}$Xe frequency, $\kappa$.
%, corresponding to the experimental conditions of Ref.~\cite{lim18} ($T=120$~C).
}
%\vspace{0.5cm}
\label{tbl-kappa}
\begin{tabular}{lrr} 
\hline\hline
Quantity & This work & Experiment \\ \hline
$J_{1}$ ($10^{-27}$~Hz\,m$^{3}$) & $2.24\pm{^{0.10}_{0.21}}\,{^a}$ & --\\ 
%$J_{^{129}{\rm Xe}-{^3}{\rm He}}$ (Hz) & 0.0780255$^a$ & \\
%$J_{^{129}{\rm Xe}-{^3}{\rm He}}$ (Hz) & 0.0907538$^b$ & \\
$\kappa$ & $-0.0105\pm{^{0.0005}_{0.0010}}\,{^a}$ & $-0.011\pm 0.001$$^b$\\ && $-0.009\pm 0.0004$$^c$\\
\hline\hline
\end{tabular}
\begin{flushleft}
{\footnotesize
%$^a$~Calculated using $^3$He number density obtained from the ideal gas law $n_{\rm He}=p/kT = 3.439\times 10^{25}$~m$^{-3}$.\\
%$^b$~Using $n_{\rm He} = 4\times 10^{25}$ m$^{-3}$.
$^a$~Error margins due to the choice of $V\!\left(R\right)$ and $J\!\left(R\right)$ are given as super- and subscripts, respectively (see text). At $T = 120$~C.\\
$^b$~Ref.~\onlinecite{lim18}. Measurement at 120~C.\\
$^c$~Ref.~\onlinecite{ter18}. The corresponding $^3$He shift measurements give $\kappa = -0.007\pm 0.001$. At $T=28$~C~\cite{ter19}.
}
\end{flushleft}
\end{table}

At the experimental conditions of Ref.~\onlinecite{lim18}, $T=120$~C and $n_{\rm He}=4\times 10^{19}$~cm$^{-3}$, the presently calculated $J_1$ corresponds to $J_{\rm XeHe}=0.09$~Hz, retaining the first term of Eq.~(\ref{eqn-virial}). The resulting $^{129}$Xe frequency shift is orders of magnitude larger than what can be expected from novel physical phenomenology, in the nHz range~\cite{saf18}. This underlines the necessity of eliminating the influence of SSC in co-magnetometer investigations. At the same time, such couplings are large enough to offer a potential tool to investigate weakly bound systems via NMR techniques.

\begin{figure}[t!]
\includegraphics[width=0.9\columnwidth]{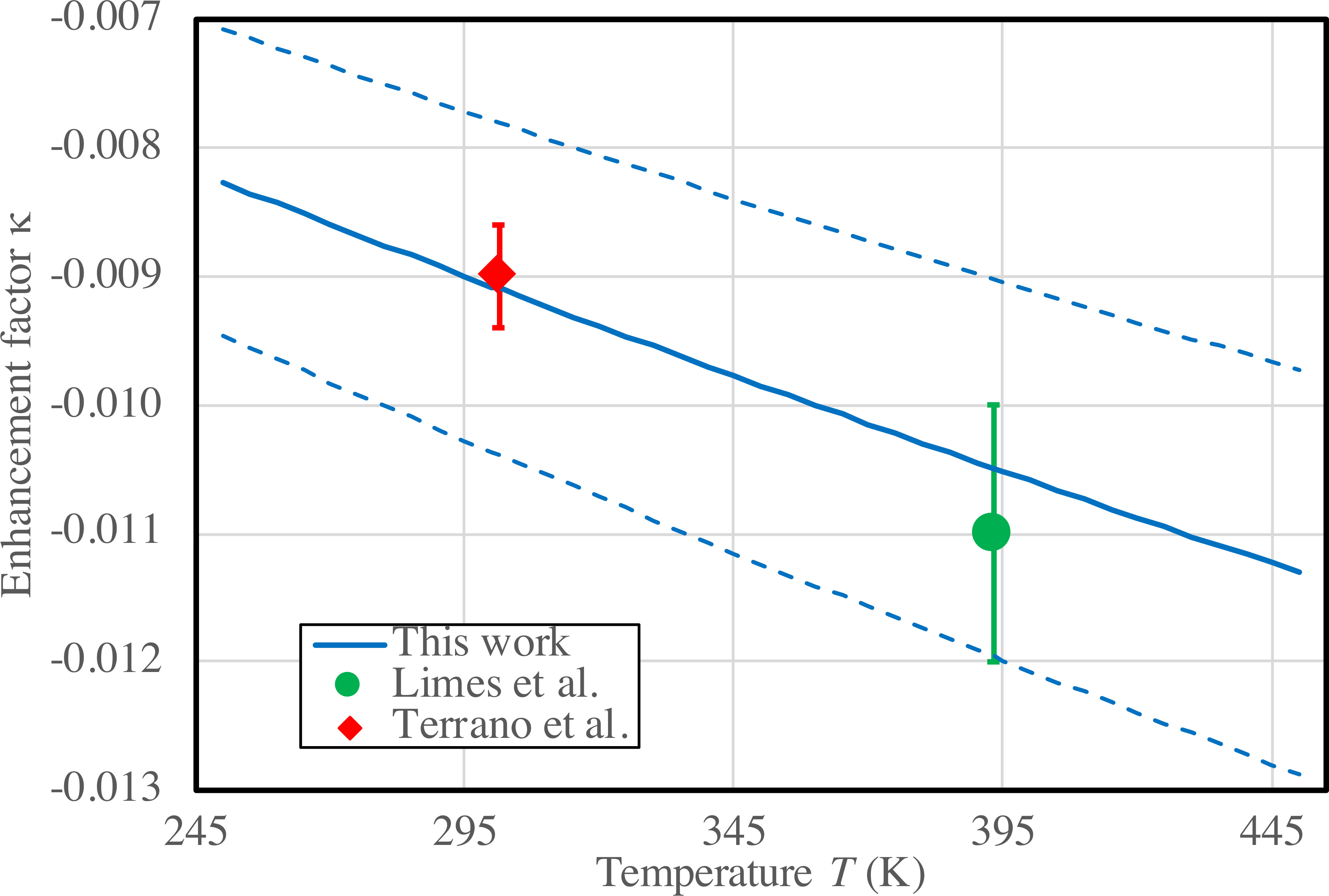}
\caption{Calculated temperature dependence of the enhancement factor $\kappa$  of the $^{129}$Xe frequency in the presence of $^3$He. Experimental data by Limes {\em et al}.~\cite{lim18} and Terrano {\em et al.}~\cite{ter18} are also given. %The measurement temperature of the latter experiment was not quoted, hence the datum is represented by a horizontal line. 
The error margins of all results are  shown.}
\label{fig-kappa}
\end{figure}

Fig.~\ref{fig-kappa} shows the calculated temperature dependence of $\kappa$ as well as the existing experimental results. In  qualitative agreement with Ref.~\onlinecite{lim18}, $\kappa$ is very modestly temperature dependent. In the depicted temperature range,  $\kappa$ can be fitted to linear dependence $\kappa = \kappa_{0} + \kappa_{1}T$ with $\kappa_{0} = -0.00453$ and $\kappa_{1}=-1.51\times 10^{-5}$~K$^{-1}$.
%\begin{figure}[htb!]
%\includegraphics[width=0.9\columnwidth]{j1.eps}
%\caption{Calculated temperature (in K) dependence of the second virial coefficient $J_{1}\!\left(^{129}{\rm Xe}-{^3}{\rm He}\right)$ (in Hz\,m$^3$) of the spin-spin coupling constant between $^{129}$Xe and $^3$He nuclei.}
%\end{figure}

Similar methodology can be used to evaluate $\kappa$ for other noble-gas pairs, {\em e.g.}, $^{129}$Xe-$^{131}$Xe. We can use data from first-principles calculations of the corresponding $J\!\left(R\right)$ and $V\!\left(R\right)$ curves in Refs.~\onlinecite{vaa13} [Eq.~(7) of that paper] and \onlinecite{hanni_2004}, respectively, in Eqs.~(\ref{eqn-j1}) and (\ref{eqn-kappa}), with $\gamma_{^{131}{\rm Xe}}$ replacing $\gamma_{^3{\rm He}}$ in the denominator of the latter equation. This leads to $\kappa = -0.352, -0.347$, and $-0.342$ at $T = 80$, 100, and 120~C, for the indirect $^{129}$Xe-$^{131}$Xe SSC. 

\section{Conclusion}

%In conclusion, w
We have computationally investigated the frequency shift of $^{129}$Xe due to the indirect spin-spin coupling to $^3$He in low-pressure gas, corresponding to recent co-magnetometer experiments. Using state-of-the-art first-principles electronic structure theory, the second virial coefficient of the SSC constant between the two isotopes was calculated and converted to the enhancement factor $\kappa$, with results that are in excellent agreement with  the 
%the experimental data from the 
two recent measurements. In reaching this quantitative agreement, both {\em ab initio\/} electron correlation treatment and consideration of relativistic influences were necessary. We trust that calculations of the present type are useful in eliminating the frequency shifts due to standard phenomenology in search of novel physics, as well as pave the way for precision determination of small interactions in materials NMR.

%\section{Conclusions}
%\noindent T\"anne tulee

\section*{Acknowledgements}
%The author is grateful to M.~V.~Romalis for suggesting the topic of this study, as well as useful discussions. 
%\begin{acknowledgement}
Academy of Finland (grant no.\ 296292) and University of Oulu (Kvantum Institute) are thanked for financial support, as well as CSC-IT Center for Science (Espoo, Finland) and Finnish Grid and Cloud Infrastructure (persistent identifier urn:nbn:fi:research-infras-2016072533), for computational resources.
%\end{acknowledgement}

%\begin{suppinfo}
%The temperature
%                  dependence of the second virial coefficient of Xe
%                  hyperfine coupling, for the basis-set used to obtain
%                  the Rb-Xe PEC, and for the fitting details of the
%                  calculated potential energy curves.
%\end{suppinfo}

%\input{bibliography.bbl}
\bibliography{bibliography}

\begin{thebibliography}{61}%
\makeatletter
\providecommand \@ifxundefined [1]{%
 \@ifx{#1\undefined}
}%
\providecommand \@ifnum [1]{%
 \ifnum #1\expandafter \@firstoftwo
 \else \expandafter \@secondoftwo
 \fi
}%
\providecommand \@ifx [1]{%
 \ifx #1\expandafter \@firstoftwo
 \else \expandafter \@secondoftwo
 \fi
}%
\providecommand \natexlab [1]{#1}%
\providecommand \enquote  [1]{``#1''}%
\providecommand \bibnamefont  [1]{#1}%
\providecommand \bibfnamefont [1]{#1}%
\providecommand \citenamefont [1]{#1}%
\providecommand \href@noop [0]{\@secondoftwo}%
\providecommand \href [0]{\begingroup \@sanitize@url \@href}%
\providecommand \@href[1]{\@@startlink{#1}\@@href}%
\providecommand \@@href[1]{\endgroup#1\@@endlink}%
\providecommand \@sanitize@url [0]{\catcode `\\12\catcode `\$12\catcode
  `\&12\catcode `\#12\catcode `\^12\catcode `\_12\catcode `\%12\relax}%
\providecommand \@@startlink[1]{}%
\providecommand \@@endlink[0]{}%
\providecommand \url  [0]{\begingroup\@sanitize@url \@url }%
\providecommand \@url [1]{\endgroup\@href {#1}{\urlprefix }}%
\providecommand \urlprefix  [0]{URL }%
\providecommand \Eprint [0]{\href }%
\providecommand \doibase [0]{http://dx.doi.org/}%
\providecommand \selectlanguage [0]{\@gobble}%
\providecommand \bibinfo  [0]{\@secondoftwo}%
\providecommand \bibfield  [0]{\@secondoftwo}%
\providecommand \translation [1]{[#1]}%
\providecommand \BibitemOpen [0]{}%
\providecommand \bibitemStop [0]{}%
\providecommand \bibitemNoStop [0]{.\EOS\space}%
\providecommand \EOS [0]{\spacefactor3000\relax}%
\providecommand \BibitemShut  [1]{\csname bibitem#1\endcsname}%
\let\auto@bib@innerbib\@empty
%</preamble>
\bibitem [{\citenamefont {Levitt}(2001)}]{lev01}%
  \BibitemOpen
  \bibfield  {author} {\bibinfo {author} {\bibfnamefont {M.~H.}\ \bibnamefont
  {Levitt}},\ }\href@noop {} {\emph {\bibinfo {title} {Spin Dynamics}}}\
  (\bibinfo  {publisher} {Wiley},\ \bibinfo {year} {2001})\BibitemShut
  {NoStop}%
\bibitem [{\citenamefont {Helgaker}, \citenamefont {Jaszu\'{n}ski},\ and\
  \citenamefont {Ruud}(1999)}]{helgaker_ab_1999}%
  \BibitemOpen
  \bibfield  {author} {\bibinfo {author} {\bibfnamefont {T.}~\bibnamefont
  {Helgaker}}, \bibinfo {author} {\bibfnamefont {M.}~\bibnamefont
  {Jaszu\'{n}ski}}, \ and\ \bibinfo {author} {\bibfnamefont {K.}~\bibnamefont
  {Ruud}},\ }\href
  {http://pubs3.acs.org/acs/journals/doilookup?in\_doi=10.1021/cr960017t}
  {\bibfield  {journal} {\bibinfo  {journal} {Chem. Rev.}\ }\textbf {\bibinfo
  {volume} {99}},\ \bibinfo {pages} {293} (\bibinfo {year} {1999})}\BibitemShut
  {NoStop}%
\bibitem [{\citenamefont {Salsbury}\ and\ \citenamefont
  {Harris}(1998)}]{sal98}%
  \BibitemOpen
  \bibfield  {author} {\bibinfo {author} {\bibfnamefont {F.~R.}\ \bibnamefont
  {Salsbury}}\ and\ \bibinfo {author} {\bibfnamefont {R.~A.}\ \bibnamefont
  {Harris}},\ }\href@noop {} {\bibfield  {journal} {\bibinfo  {journal} {Mol.
  Phys.}\ }\textbf {\bibinfo {volume} {94}},\ \bibinfo {pages} {307} (\bibinfo
  {year} {1998})}\BibitemShut {NoStop}%
\bibitem [{\citenamefont {Pecul}(2000)}]{pec00}%
  \BibitemOpen
  \bibfield  {author} {\bibinfo {author} {\bibfnamefont {M.}~\bibnamefont
  {Pecul}},\ }\href@noop {} {\bibfield  {journal} {\bibinfo  {journal} {J.
  Chem. Phys.}\ }\textbf {\bibinfo {volume} {113}},\ \bibinfo {pages} {10835}
  (\bibinfo {year} {2000})}\BibitemShut {NoStop}%
\bibitem [{\citenamefont {Pecul}, \citenamefont {Sadlej},\ and\ \citenamefont
  {Leszczynski}(2001)}]{pec01}%
  \BibitemOpen
  \bibfield  {author} {\bibinfo {author} {\bibfnamefont {M.}~\bibnamefont
  {Pecul}}, \bibinfo {author} {\bibfnamefont {J.}~\bibnamefont {Sadlej}}, \
  and\ \bibinfo {author} {\bibfnamefont {J.}~\bibnamefont {Leszczynski}},\
  }\href@noop {} {\bibfield  {journal} {\bibinfo  {journal} {J. Chem. Phys.}\
  }\textbf {\bibinfo {volume} {115}},\ \bibinfo {pages} {5498} (\bibinfo {year}
  {2001})}\BibitemShut {NoStop}%
\bibitem [{\citenamefont {Bagno}, \citenamefont {Saielli},\ and\ \citenamefont
  {Scorrano}(2001)}]{bag01}%
  \BibitemOpen
  \bibfield  {author} {\bibinfo {author} {\bibfnamefont {A.}~\bibnamefont
  {Bagno}}, \bibinfo {author} {\bibfnamefont {G.}~\bibnamefont {Saielli}}, \
  and\ \bibinfo {author} {\bibfnamefont {G.}~\bibnamefont {Scorrano}},\
  }\href@noop {} {\bibfield  {journal} {\bibinfo  {journal} {Angew. Chem. Int.
  Ed.}\ }\textbf {\bibinfo {volume} {40}},\ \bibinfo {pages} {2532} (\bibinfo
  {year} {2001})}\BibitemShut {NoStop}%
\bibitem [{\citenamefont {Bagno}, \citenamefont {Saielli},\ and\ \citenamefont
  {Scorrano}(2002)}]{bag02}%
  \BibitemOpen
  \bibfield  {author} {\bibinfo {author} {\bibfnamefont {A.}~\bibnamefont
  {Bagno}}, \bibinfo {author} {\bibfnamefont {G.}~\bibnamefont {Saielli}}, \
  and\ \bibinfo {author} {\bibfnamefont {G.}~\bibnamefont {Scorrano}},\
  }\href@noop {} {\bibfield  {journal} {\bibinfo  {journal} {Chem.-Eur. J.}\
  }\textbf {\bibinfo {volume} {8}},\ \bibinfo {pages} {2047} (\bibinfo {year}
  {2002})}\BibitemShut {NoStop}%
\bibitem [{\citenamefont {Vaara}, \citenamefont {Hanni},\ and\ \citenamefont
  {Jokisaari}(2013)}]{vaa13}%
  \BibitemOpen
  \bibfield  {author} {\bibinfo {author} {\bibfnamefont {J.}~\bibnamefont
  {Vaara}}, \bibinfo {author} {\bibfnamefont {M.}~\bibnamefont {Hanni}}, \ and\
  \bibinfo {author} {\bibfnamefont {J.}~\bibnamefont {Jokisaari}},\ }\href@noop
  {} {\bibfield  {journal} {\bibinfo  {journal} {J. Chem. Phys.}\ }\textbf
  {\bibinfo {volume} {138}},\ \bibinfo {pages} {104313} (\bibinfo {year}
  {2013})}\BibitemShut {NoStop}%
\bibitem [{\citenamefont {Ledbetter}\ \emph {et~al.}(2012)\citenamefont
  {Ledbetter}, \citenamefont {Saielli}, \citenamefont {Bagno}, \citenamefont
  {Tran},\ and\ \citenamefont {Romalis}}]{led12}%
  \BibitemOpen
  \bibfield  {author} {\bibinfo {author} {\bibfnamefont {M.~P.}\ \bibnamefont
  {Ledbetter}}, \bibinfo {author} {\bibfnamefont {G.}~\bibnamefont {Saielli}},
  \bibinfo {author} {\bibfnamefont {A.}~\bibnamefont {Bagno}}, \bibinfo
  {author} {\bibfnamefont {N.}~\bibnamefont {Tran}}, \ and\ \bibinfo {author}
  {\bibfnamefont {M.~V.}\ \bibnamefont {Romalis}},\ }\href@noop {} {\bibfield
  {journal} {\bibinfo  {journal} {Proc. Nat. Acad. Sci.}\ }\textbf {\bibinfo
  {volume} {109}},\ \bibinfo {pages} {12393} (\bibinfo {year}
  {2012})}\BibitemShut {NoStop}%
\bibitem [{\citenamefont {Jokisaari}\ and\ \citenamefont
  {Vaara}(2013)}]{jok13}%
  \BibitemOpen
  \bibfield  {author} {\bibinfo {author} {\bibfnamefont {J.}~\bibnamefont
  {Jokisaari}}\ and\ \bibinfo {author} {\bibfnamefont {J.}~\bibnamefont
  {Vaara}},\ }\href@noop {} {\bibfield  {journal} {\bibinfo  {journal} {Phys.
  Chem. Chem. Phys.}\ }\textbf {\bibinfo {volume} {15}},\ \bibinfo {pages}
  {11427} (\bibinfo {year} {2013})}\BibitemShut {NoStop}%
\bibitem [{lim()}]{lim18}%
  \BibitemOpen
  \href@noop {} {}\bibinfo {note} {M.~E.~Limes, N.~Dural, M.~V.~Romalis,
  E.~L.~Foley, T.~W.~Kornack, A.~Nelson, and L.~R.~Grisham, arXiv:1805.11578
  [physics.atom-ph] (2018)}\BibitemShut {NoStop}%
\bibitem [{ter()}]{ter18}%
  \BibitemOpen
  \href@noop {} {}\bibinfo {note} {{W. A. Terrano, J. Meinel, N. Sachdeva, T.
  Chupp, D. Degenkolb, P. Fierlinger, F. Kuchler, and J. T. Singh,
  arXiv:1807.11119 [physics.atom-ph] (2018)}}\BibitemShut {NoStop}%
\bibitem [{\citenamefont {Walker}(1989)}]{walker_1989}%
  \BibitemOpen
  \bibfield  {author} {\bibinfo {author} {\bibfnamefont {T.~G.}\ \bibnamefont
  {Walker}},\ }\href@noop {} {\bibfield  {journal} {\bibinfo  {journal} {Phys.
  Rev. A}\ }\textbf {\bibinfo {volume} {40}},\ \bibinfo {pages} {4959}
  (\bibinfo {year} {1989})}\BibitemShut {NoStop}%
\bibitem [{\citenamefont {Hanni}\ \emph {et~al.}(2017)\citenamefont {Hanni},
  \citenamefont {Lantto}, \citenamefont {Repisk\'y}, \citenamefont {Mare\v{s}},
  \citenamefont {Saam},\ and\ \citenamefont {Vaara}}]{han17}%
  \BibitemOpen
  \bibfield  {author} {\bibinfo {author} {\bibfnamefont {M.}~\bibnamefont
  {Hanni}}, \bibinfo {author} {\bibfnamefont {P.}~\bibnamefont {Lantto}},
  \bibinfo {author} {\bibfnamefont {M.}~\bibnamefont {Repisk\'y}}, \bibinfo
  {author} {\bibfnamefont {J.}~\bibnamefont {Mare\v{s}}}, \bibinfo {author}
  {\bibfnamefont {B.}~\bibnamefont {Saam}}, \ and\ \bibinfo {author}
  {\bibfnamefont {J.}~\bibnamefont {Vaara}},\ }\href@noop {} {\bibfield
  {journal} {\bibinfo  {journal} {Phys. Rev. A}\ }\textbf {\bibinfo {volume}
  {95}},\ \bibinfo {pages} {032509} (\bibinfo {year} {2017})}\BibitemShut
  {NoStop}%
\bibitem [{\citenamefont {Walker}\ and\ \citenamefont
  {Happer}(1997)}]{walker_1997}%
  \BibitemOpen
  \bibfield  {author} {\bibinfo {author} {\bibfnamefont {T.~G.}\ \bibnamefont
  {Walker}}\ and\ \bibinfo {author} {\bibfnamefont {W.}~\bibnamefont
  {Happer}},\ }\href@noop {} {\bibfield  {journal} {\bibinfo  {journal} {Rev.
  Mod. Phys.}\ }\textbf {\bibinfo {volume} {69}},\ \bibinfo {pages} {629}
  (\bibinfo {year} {1997})}\BibitemShut {NoStop}%
\bibitem [{\citenamefont {Safronova}\ \emph {et~al.}(2018)\citenamefont
  {Safronova}, \citenamefont {Budker}, \citenamefont {DeMille}, \citenamefont
  {Kimball}, \citenamefont {Derevianko},\ and\ \citenamefont {Clark}}]{saf18}%
  \BibitemOpen
  \bibfield  {author} {\bibinfo {author} {\bibfnamefont {M.~S.}\ \bibnamefont
  {Safronova}}, \bibinfo {author} {\bibfnamefont {D.}~\bibnamefont {Budker}},
  \bibinfo {author} {\bibfnamefont {D.}~\bibnamefont {DeMille}}, \bibinfo
  {author} {\bibfnamefont {D.~F.~J.}\ \bibnamefont {Kimball}}, \bibinfo
  {author} {\bibfnamefont {A.}~\bibnamefont {Derevianko}}, \ and\ \bibinfo
  {author} {\bibfnamefont {C.~W.}\ \bibnamefont {Clark}},\ }\href@noop {}
  {\bibfield  {journal} {\bibinfo  {journal} {Rev. Mod. Phys.}\ }\textbf
  {\bibinfo {volume} {90}},\ \bibinfo {pages} {025008} (\bibinfo {year}
  {2018})}\BibitemShut {NoStop}%
\bibitem [{\citenamefont {Graham}\ \emph {et~al.}(2018)\citenamefont {Graham},
  \citenamefont {Kaplan}, \citenamefont {Mardon}, \citenamefont {Rajendran},
  \citenamefont {Terrano}, \citenamefont {Trahms},\ and\ \citenamefont
  {Wilkason}}]{gra18}%
  \BibitemOpen
  \bibfield  {author} {\bibinfo {author} {\bibfnamefont {P.~W.}\ \bibnamefont
  {Graham}}, \bibinfo {author} {\bibfnamefont {D.~E.}\ \bibnamefont {Kaplan}},
  \bibinfo {author} {\bibfnamefont {J.}~\bibnamefont {Mardon}}, \bibinfo
  {author} {\bibfnamefont {S.}~\bibnamefont {Rajendran}}, \bibinfo {author}
  {\bibfnamefont {W.~A.}\ \bibnamefont {Terrano}}, \bibinfo {author}
  {\bibfnamefont {L.}~\bibnamefont {Trahms}}, \ and\ \bibinfo {author}
  {\bibfnamefont {T.}~\bibnamefont {Wilkason}},\ }\href {\doibase
  10.1103/PhysRevD.97.055006} {\bibfield  {journal} {\bibinfo  {journal} {Phys.
  Rev. D}\ }\textbf {\bibinfo {volume} {97}},\ \bibinfo {pages} {055006}
  (\bibinfo {year} {2018})}\BibitemShut {NoStop}%
\bibitem [{\citenamefont {Glenday}\ \emph {et~al.}(2008)\citenamefont
  {Glenday}, \citenamefont {Cramer}, \citenamefont {Phillips},\ and\
  \citenamefont {Walsworth}}]{gle08}%
  \BibitemOpen
  \bibfield  {author} {\bibinfo {author} {\bibfnamefont {A.~G.}\ \bibnamefont
  {Glenday}}, \bibinfo {author} {\bibfnamefont {C.~E.}\ \bibnamefont {Cramer}},
  \bibinfo {author} {\bibfnamefont {D.~F.}\ \bibnamefont {Phillips}}, \ and\
  \bibinfo {author} {\bibfnamefont {R.~L.}\ \bibnamefont {Walsworth}},\
  }\href@noop {} {\bibfield  {journal} {\bibinfo  {journal} {Phys. Rev. Lett.}\
  }\textbf {\bibinfo {volume} {101}},\ \bibinfo {pages} {261801} (\bibinfo
  {year} {2008})}\BibitemShut {NoStop}%
\bibitem [{\citenamefont {Bulatowicz}\ \emph {et~al.}(2013)\citenamefont
  {Bulatowicz}, \citenamefont {Griffith}, \citenamefont {Larsen}, \citenamefont
  {Mirijanian}, \citenamefont {Fu}, \citenamefont {Smith}, \citenamefont
  {Snow}, \citenamefont {Yan},\ and\ \citenamefont {Walker}}]{bul13}%
  \BibitemOpen
  \bibfield  {author} {\bibinfo {author} {\bibfnamefont {M.}~\bibnamefont
  {Bulatowicz}}, \bibinfo {author} {\bibfnamefont {R.}~\bibnamefont
  {Griffith}}, \bibinfo {author} {\bibfnamefont {M.}~\bibnamefont {Larsen}},
  \bibinfo {author} {\bibfnamefont {J.}~\bibnamefont {Mirijanian}}, \bibinfo
  {author} {\bibfnamefont {C.~B.}\ \bibnamefont {Fu}}, \bibinfo {author}
  {\bibfnamefont {E.}~\bibnamefont {Smith}}, \bibinfo {author} {\bibfnamefont
  {W.~M.}\ \bibnamefont {Snow}}, \bibinfo {author} {\bibfnamefont
  {H.}~\bibnamefont {Yan}}, \ and\ \bibinfo {author} {\bibfnamefont {T.~G.}\
  \bibnamefont {Walker}},\ }\href@noop {} {\bibfield  {journal} {\bibinfo
  {journal} {Phys. Rev. Lett.}\ }\textbf {\bibinfo {volume} {111}},\ \bibinfo
  {pages} {102001} (\bibinfo {year} {2013})}\BibitemShut {NoStop}%
\bibitem [{\citenamefont {Tullney}\ \emph {et~al.}(2013)\citenamefont
  {Tullney}, \citenamefont {Allmendiger}, \citenamefont {Burghoff},
  \citenamefont {Heil}, \citenamefont {Karpuk}, \citenamefont {Kilian},
  \citenamefont {Knappe-Gr\"uneberg}, \citenamefont {M\"uller}, \citenamefont
  {Schmidt}, \citenamefont {Schnabel}, \citenamefont {Seifert}, \citenamefont
  {Sobolev},\ and\ \citenamefont {Trahms}}]{tul13}%
  \BibitemOpen
  \bibfield  {author} {\bibinfo {author} {\bibfnamefont {K.}~\bibnamefont
  {Tullney}}, \bibinfo {author} {\bibfnamefont {F.}~\bibnamefont
  {Allmendiger}}, \bibinfo {author} {\bibfnamefont {M.}~\bibnamefont
  {Burghoff}}, \bibinfo {author} {\bibfnamefont {W.}~\bibnamefont {Heil}},
  \bibinfo {author} {\bibfnamefont {S.}~\bibnamefont {Karpuk}}, \bibinfo
  {author} {\bibfnamefont {W.}~\bibnamefont {Kilian}}, \bibinfo {author}
  {\bibfnamefont {S.}~\bibnamefont {Knappe-Gr\"uneberg}}, \bibinfo {author}
  {\bibfnamefont {W.}~\bibnamefont {M\"uller}}, \bibinfo {author}
  {\bibfnamefont {U.}~\bibnamefont {Schmidt}}, \bibinfo {author} {\bibfnamefont
  {A.}~\bibnamefont {Schnabel}}, \bibinfo {author} {\bibfnamefont
  {F.}~\bibnamefont {Seifert}}, \bibinfo {author} {\bibfnamefont
  {Y.}~\bibnamefont {Sobolev}}, \ and\ \bibinfo {author} {\bibfnamefont
  {L.}~\bibnamefont {Trahms}},\ }\href@noop {} {\bibfield  {journal} {\bibinfo
  {journal} {Phys. Rev. Lett.}\ }\textbf {\bibinfo {volume} {111}},\ \bibinfo
  {pages} {100801} (\bibinfo {year} {2013})}\BibitemShut {NoStop}%
\bibitem [{\citenamefont {Rosenberry}\ and\ \citenamefont
  {Chupp}(2001)}]{ros01}%
  \BibitemOpen
  \bibfield  {author} {\bibinfo {author} {\bibfnamefont {M.~A.}\ \bibnamefont
  {Rosenberry}}\ and\ \bibinfo {author} {\bibfnamefont {T.~E.}\ \bibnamefont
  {Chupp}},\ }\href@noop {} {\bibfield  {journal} {\bibinfo  {journal} {Phys.
  Rev. Lett.}\ }\textbf {\bibinfo {volume} {86}},\ \bibinfo {pages} {22}
  (\bibinfo {year} {2001})}\BibitemShut {NoStop}%
\bibitem [{\citenamefont {Bear}\ \emph {et~al.}(2000)\citenamefont {Bear},
  \citenamefont {Stoner}, \citenamefont {Walsworth}, \citenamefont
  {Kosteleck\'y},\ and\ \citenamefont {Lane}}]{bea00}%
  \BibitemOpen
  \bibfield  {author} {\bibinfo {author} {\bibfnamefont {D.}~\bibnamefont
  {Bear}}, \bibinfo {author} {\bibfnamefont {R.~E.}\ \bibnamefont {Stoner}},
  \bibinfo {author} {\bibfnamefont {R.~L.}\ \bibnamefont {Walsworth}}, \bibinfo
  {author} {\bibfnamefont {V.~A.}\ \bibnamefont {Kosteleck\'y}}, \ and\
  \bibinfo {author} {\bibfnamefont {C.~D.}\ \bibnamefont {Lane}},\ }\href@noop
  {} {\bibfield  {journal} {\bibinfo  {journal} {Phys. Rev. Lett.}\ }\textbf
  {\bibinfo {volume} {85}},\ \bibinfo {pages} {5038} (\bibinfo {year}
  {2000})}\BibitemShut {NoStop}%
\bibitem [{\citenamefont {Heckman}, \citenamefont {Ledbetter},\ and\
  \citenamefont {Romalis}(2003)}]{hec03}%
  \BibitemOpen
  \bibfield  {author} {\bibinfo {author} {\bibfnamefont {J.~J.}\ \bibnamefont
  {Heckman}}, \bibinfo {author} {\bibfnamefont {M.~P.}\ \bibnamefont
  {Ledbetter}}, \ and\ \bibinfo {author} {\bibfnamefont {M.~V.}\ \bibnamefont
  {Romalis}},\ }\href@noop {} {\bibfield  {journal} {\bibinfo  {journal} {Phys.
  Rev. Lett.}\ }\textbf {\bibinfo {volume} {91}},\ \bibinfo {pages} {067601}
  (\bibinfo {year} {2003})}\BibitemShut {NoStop}%
\bibitem [{\citenamefont {Reiher}\ and\ \citenamefont {Wolf}(2009)}]{rei09}%
  \BibitemOpen
  \bibfield  {author} {\bibinfo {author} {\bibfnamefont {M.}~\bibnamefont
  {Reiher}}\ and\ \bibinfo {author} {\bibfnamefont {A.}~\bibnamefont {Wolf}},\
  }\href@noop {} {\emph {\bibinfo {title} {Relativistic Quantum Chemistry}}}\
  (\bibinfo  {publisher} {Wiley-VCH},\ \bibinfo {year} {2009})\BibitemShut
  {NoStop}%
\bibitem [{\citenamefont {Moss}(1973)}]{mos73}%
  \BibitemOpen
  \bibfield  {author} {\bibinfo {author} {\bibfnamefont {R.~E.}\ \bibnamefont
  {Moss}},\ }\href@noop {} {\emph {\bibinfo {title} {Advanced Molecular Quantum
  Mechanics}}}\ (\bibinfo  {publisher} {Chapman and Hall},\ \bibinfo {year}
  {1973})\BibitemShut {NoStop}%
\bibitem [{\citenamefont {Buckingham}\ and\ \citenamefont
  {Pople}(1956)}]{buckingham_1956}%
  \BibitemOpen
  \bibfield  {author} {\bibinfo {author} {\bibfnamefont {A.~D.}\ \bibnamefont
  {Buckingham}}\ and\ \bibinfo {author} {\bibfnamefont {J.~A.}\ \bibnamefont
  {Pople}},\ }\href@noop {} {\bibfield  {journal} {\bibinfo  {journal}
  {Discuss. Faraday Soc.}\ }\textbf {\bibinfo {volume} {22}},\ \bibinfo {pages}
  {17} (\bibinfo {year} {1956})}\BibitemShut {NoStop}%
\bibitem [{\citenamefont {Schaefer}\ \emph {et~al.}(1989)\citenamefont
  {Schaefer}, \citenamefont {Cates}, \citenamefont {Chien}, \citenamefont
  {Gonatas}, \citenamefont {Happer},\ and\ \citenamefont
  {Walker}}]{schaefer_1989}%
  \BibitemOpen
  \bibfield  {author} {\bibinfo {author} {\bibfnamefont {S.~R.}\ \bibnamefont
  {Schaefer}}, \bibinfo {author} {\bibfnamefont {G.~D.}\ \bibnamefont {Cates}},
  \bibinfo {author} {\bibfnamefont {T.-R.}\ \bibnamefont {Chien}}, \bibinfo
  {author} {\bibfnamefont {D.}~\bibnamefont {Gonatas}}, \bibinfo {author}
  {\bibfnamefont {W.}~\bibnamefont {Happer}}, \ and\ \bibinfo {author}
  {\bibfnamefont {T.~G.}\ \bibnamefont {Walker}},\ }\href@noop {} {\bibfield
  {journal} {\bibinfo  {journal} {Phys. Rev. A}\ }\textbf {\bibinfo {volume}
  {39}},\ \bibinfo {pages} {5613} (\bibinfo {year} {1989})}\BibitemShut
  {NoStop}%
\bibitem [{\citenamefont {Werner}\ \emph {et~al.}(2012)\citenamefont {Werner},
  \citenamefont {Knowles}, \citenamefont {Knizia}, \citenamefont {Manby},\ and\
  \citenamefont {Sch{\"u}tz}}]{wer12}%
  \BibitemOpen
  \bibfield  {author} {\bibinfo {author} {\bibfnamefont {H.-J.}\ \bibnamefont
  {Werner}}, \bibinfo {author} {\bibfnamefont {P.~J.}\ \bibnamefont {Knowles}},
  \bibinfo {author} {\bibfnamefont {G.}~\bibnamefont {Knizia}}, \bibinfo
  {author} {\bibfnamefont {F.~R.}\ \bibnamefont {Manby}}, \ and\ \bibinfo
  {author} {\bibfnamefont {M.}~\bibnamefont {Sch{\"u}tz}},\ }\href@noop {}
  {\bibfield  {journal} {\bibinfo  {journal} {WIREs Comput. Mol. Sci.}\
  }\textbf {\bibinfo {volume} {2}},\ \bibinfo {pages} {242} (\bibinfo {year}
  {2012})}\BibitemShut {NoStop}%
\bibitem [{mol()}]{molpro}%
  \BibitemOpen
  \href@noop {} {}\bibinfo {note} {{\rm {\sc Molpro}, version 2015.1, a package
  of ab initio programs, H.-J.~Werner, P.~J.~Knowles, G.~Knizia, F.~R.~Manby,
  M.~Sch\"utz {\em et al.}, see {\tt http://www.molpro.net}.}}\BibitemShut
  {Stop}%
\bibitem [{\citenamefont {Dunning~Jr.}(1989)}]{dun89}%
  \BibitemOpen
  \bibfield  {author} {\bibinfo {author} {\bibfnamefont {T.~H.}\ \bibnamefont
  {Dunning~Jr.}},\ }\href@noop {} {\bibfield  {journal} {\bibinfo  {journal}
  {J. Chem. Phys.}\ }\textbf {\bibinfo {volume} {90}},\ \bibinfo {pages} {1007}
  (\bibinfo {year} {1989})}\BibitemShut {NoStop}%
\bibitem [{\citenamefont {Peterson}\ \emph {et~al.}(2003)\citenamefont
  {Peterson}, \citenamefont {Figgen}, \citenamefont {Goll}, \citenamefont
  {Stoll},\ and\ \citenamefont {Dolg}}]{peterson_2003}%
  \BibitemOpen
  \bibfield  {author} {\bibinfo {author} {\bibfnamefont {K.~A.}\ \bibnamefont
  {Peterson}}, \bibinfo {author} {\bibfnamefont {D.}~\bibnamefont {Figgen}},
  \bibinfo {author} {\bibfnamefont {E.}~\bibnamefont {Goll}}, \bibinfo {author}
  {\bibfnamefont {H.}~\bibnamefont {Stoll}}, \ and\ \bibinfo {author}
  {\bibfnamefont {M.}~\bibnamefont {Dolg}},\ }\href@noop {} {\bibfield
  {journal} {\bibinfo  {journal} {J. Chem. Phys.}\ }\textbf {\bibinfo {volume}
  {119}},\ \bibinfo {pages} {11113} (\bibinfo {year} {2003})}\BibitemShut
  {NoStop}%
\bibitem [{\citenamefont {Slav{\'i}{\v c}ek}\ \emph {et~al.}(2003)\citenamefont
  {Slav{\'i}{\v c}ek}, \citenamefont {Kalus}, \citenamefont {Pa{\v s}ka},
  \citenamefont {Odv{\'a}rkov{\'a}}, \citenamefont {Hobza},\ and\ \citenamefont
  {Malijevsk{\'y}}}]{slavicek_2003}%
  \BibitemOpen
  \bibfield  {author} {\bibinfo {author} {\bibfnamefont {P.}~\bibnamefont
  {Slav{\'i}{\v c}ek}}, \bibinfo {author} {\bibfnamefont {R.}~\bibnamefont
  {Kalus}}, \bibinfo {author} {\bibfnamefont {P.}~\bibnamefont {Pa{\v s}ka}},
  \bibinfo {author} {\bibfnamefont {I.}~\bibnamefont {Odv{\'a}rkov{\'a}}},
  \bibinfo {author} {\bibfnamefont {P.}~\bibnamefont {Hobza}}, \ and\ \bibinfo
  {author} {\bibfnamefont {A.}~\bibnamefont {Malijevsk{\'y}}},\ }\href@noop {}
  {\bibfield  {journal} {\bibinfo  {journal} {J. Chem. Phys.}\ }\textbf
  {\bibinfo {volume} {119}},\ \bibinfo {pages} {2102} (\bibinfo {year}
  {2003})}\BibitemShut {NoStop}%
\bibitem [{\citenamefont {Boys}\ and\ \citenamefont
  {Bernardi}(1970)}]{boys_1970}%
  \BibitemOpen
  \bibfield  {author} {\bibinfo {author} {\bibfnamefont {S.~F.}\ \bibnamefont
  {Boys}}\ and\ \bibinfo {author} {\bibfnamefont {F.}~\bibnamefont
  {Bernardi}},\ }\href@noop {} {\bibfield  {journal} {\bibinfo  {journal} {Mol.
  Phys.}\ }\textbf {\bibinfo {volume} {19}},\ \bibinfo {pages} {553} (\bibinfo
  {year} {1970})}\BibitemShut {NoStop}%
\bibitem [{\citenamefont {Aziz}\ and\ \citenamefont {Chen}(1977)}]{azi77}%
  \BibitemOpen
  \bibfield  {author} {\bibinfo {author} {\bibfnamefont {R.~A.}\ \bibnamefont
  {Aziz}}\ and\ \bibinfo {author} {\bibfnamefont {H.~H.}\ \bibnamefont
  {Chen}},\ }\href@noop {} {\bibfield  {journal} {\bibinfo  {journal} {J. Chem.
  Phys.}\ }\textbf {\bibinfo {volume} {67}},\ \bibinfo {pages} {5719} (\bibinfo
  {year} {1977})}\BibitemShut {NoStop}%
\bibitem [{sup()}]{supp}%
  \BibitemOpen
  \href@noop {} {}\bibinfo {note} {See Supplemental Material at (link will be
  added by the publisher) the formula of the HFD-B potential and fit parameters
  for both $V\!\left(R\right)$ and $J\!\left(R\right)$.}\BibitemShut {Stop}%
\bibitem [{\citenamefont {Rizzo}\ \emph {et~al.}(2002)\citenamefont {Rizzo},
  \citenamefont {H\"attig}, \citenamefont {Fern{\'a}ndez},\ and\ \citenamefont
  {Koch}}]{riz02}%
  \BibitemOpen
  \bibfield  {author} {\bibinfo {author} {\bibfnamefont {A.}~\bibnamefont
  {Rizzo}}, \bibinfo {author} {\bibfnamefont {C.}~\bibnamefont {H\"attig}},
  \bibinfo {author} {\bibfnamefont {B.}~\bibnamefont {Fern{\'a}ndez}}, \ and\
  \bibinfo {author} {\bibfnamefont {H.}~\bibnamefont {Koch}},\ }\href@noop {}
  {\bibfield  {journal} {\bibinfo  {journal} {J. Chem. Phys.}\ }\textbf
  {\bibinfo {volume} {117}},\ \bibinfo {pages} {2609} (\bibinfo {year}
  {2002})}\BibitemShut {NoStop}%
\bibitem [{\citenamefont {Aziz}\ \emph {et~al.}(1989)\citenamefont {Aziz},
  \citenamefont {Buck}, \citenamefont {J\'onsson}, \citenamefont
  {Ruiz-Su\'arez}, \citenamefont {Schmidt}, \citenamefont {Scoles},
  \citenamefont {Slaman},\ and\ \citenamefont {Xu}}]{azi89}%
  \BibitemOpen
  \bibfield  {author} {\bibinfo {author} {\bibfnamefont {R.~A.}\ \bibnamefont
  {Aziz}}, \bibinfo {author} {\bibfnamefont {U.}~\bibnamefont {Buck}}, \bibinfo
  {author} {\bibfnamefont {H.}~\bibnamefont {J\'onsson}}, \bibinfo {author}
  {\bibfnamefont {J.-C.}\ \bibnamefont {Ruiz-Su\'arez}}, \bibinfo {author}
  {\bibfnamefont {B.}~\bibnamefont {Schmidt}}, \bibinfo {author} {\bibfnamefont
  {G.}~\bibnamefont {Scoles}}, \bibinfo {author} {\bibfnamefont {M.~J.}\
  \bibnamefont {Slaman}}, \ and\ \bibinfo {author} {\bibfnamefont
  {J.}~\bibnamefont {Xu}},\ }\href@noop {} {\bibfield  {journal} {\bibinfo
  {journal} {J. Chem. Phys.}\ }\textbf {\bibinfo {volume} {91}},\ \bibinfo
  {pages} {6477} (\bibinfo {year} {1989})}\BibitemShut {NoStop}%
\bibitem [{DIR()}]{DIRAC17}%
  \BibitemOpen
  \href@noop {} {}\bibinfo {note} {{\sc Dirac}, a relativistic ab initio
  electronic structure program, Release {\sc Dirac17} (2017), written by
  L.~Visscher, H.~J.~{\relax Aa}.~Jensen, R.~Bast, and T.~Saue, with
  contributions from V.~Bakken, K.~G.~Dyall, S.~Dubillard, U.~Ekstr{\"o}m,
  E.~Eliav, T.~Enevoldsen, E.~Fa{\ss}hauer, T.~Fleig, O.~Fossgaard,
  A.~S.~P.~Gomes, E.~D.~Hedeg{\aa}rd, T.~Helgaker, J.~Henriksson,
  M.~Ilia{\v{s}}, {\relax Ch.}~R.~Jacob, S.~Knecht, S.~Komorovsk{\'y},
  O.~Kullie, J.~K.~L{\ae}rdahl, C.~V.~Larsen, Y.~S.~Lee, H.~S.~Nataraj,
  M.~K.~Nayak, P.~Norman, G.~Olejniczak, J.~Olsen, J.~M.~H.~Olsen, Y.~C.~Park,
  J.~K.~Pedersen, M.~Pernpointner, R.~di~Remigio, K.~Ruud, P.~Sa{\l}ek,
  B.~Schimmelpfennig, A.~Shee, J.~Sikkema, A.~J.~Thorvaldsen, J.~Thyssen,
  J.~van~Stralen, S.~Villaume, O.~Visser, T.~Winther, and S.~Yamamoto, (see
  \url{http://www.diracprogram.org})}\BibitemShut {NoStop}%
\bibitem [{\citenamefont {Repisk\'y}\ \emph {et~al.}(2009)\citenamefont
  {Repisk\'y}, \citenamefont {Komorovsky}, \citenamefont {Malkina},\ and\
  \citenamefont {Malkin}}]{respectJ}%
  \BibitemOpen
  \bibfield  {author} {\bibinfo {author} {\bibfnamefont {M.}~\bibnamefont
  {Repisk\'y}}, \bibinfo {author} {\bibfnamefont {S.}~\bibnamefont
  {Komorovsky}}, \bibinfo {author} {\bibfnamefont {O.~L.}\ \bibnamefont
  {Malkina}}, \ and\ \bibinfo {author} {\bibfnamefont {V.~G.}\ \bibnamefont
  {Malkin}},\ }\href@noop {} {\bibfield  {journal} {\bibinfo  {journal} {Chem.
  Phys.}\ }\textbf {\bibinfo {volume} {356}},\ \bibinfo {pages} {236} (\bibinfo
  {year} {2009})}\BibitemShut {NoStop}%
\bibitem [{\citenamefont {Vaara}(2007)}]{vaa07}%
  \BibitemOpen
  \bibfield  {author} {\bibinfo {author} {\bibfnamefont {J.}~\bibnamefont
  {Vaara}},\ }\href@noop {} {\bibfield  {journal} {\bibinfo  {journal} {Phys.
  Chem. Chem. Phys.}\ }\textbf {\bibinfo {volume} {9}},\ \bibinfo {pages}
  {5339} (\bibinfo {year} {2007})}\BibitemShut {NoStop}%
\bibitem [{CFO()}]{CFOUR}%
  \BibitemOpen
  \href@noop {} {}\bibinfo {note} {{\rm {\sc Cfour}, a quantum chemical program
  package written by J.~F.~Stanton, J.~Gauss, M.~E.~Harding, and P.~G.~Szalay,
  with contributions from A.~A.~Auer, R.~J.~Bartlett, U.~Benedikt, C.~Berger,
  D.~E.~Bernholdt, Y.~J.~Bomble, L.~Cheng, O.~Christiansen, M.~Heckert,
  O.~Heun, C.~Huber, T.-C.~Jagau, D.~Jonsson, J.~Jus{\'e}lius, K.~Klein,
  W.~J.~Lauderdale, D.~A.~Matthews, T.~Metzroth, L.~A.~M{\"u}ck, D.~P.~O'Neill,
  D.~R.~Price, E.~Prochnow, C.~Puzzarini, K.~Ruud, F.~Schiffmann,
  W.~Schwalbach, C.~Simmons, S.~Stopkowicz, A.~Tajti, J.~V{\'a}zquez, F.~Wang,
  and J.~D.~Watts, and the integral packages MOLECULE (J.~Alml{\"o}f and
  P.~R.~Taylor), PROPS (P.~R.~Taylor), ABACUS (T.~Helgaker, H.~J.~{\relax
  Aa}.~Jensen, P.~J{\o}rgensen, and J.~Olsen), and ECP routines by A.~V.~Mitin
  and C.~van W{\"u}llen (see \url{http://www.cfour.de})}}\BibitemShut {NoStop}%
\bibitem [{\citenamefont {Benedikt}, \citenamefont {Auer},\ and\ \citenamefont
  {Jensen}(2008)}]{ben08}%
  \BibitemOpen
  \bibfield  {author} {\bibinfo {author} {\bibfnamefont {U.}~\bibnamefont
  {Benedikt}}, \bibinfo {author} {\bibfnamefont {A.~A.}\ \bibnamefont {Auer}},
  \ and\ \bibinfo {author} {\bibfnamefont {F.}~\bibnamefont {Jensen}},\
  }\href@noop {} {\bibfield  {journal} {\bibinfo  {journal} {J. Chem. Phys.}\
  }\textbf {\bibinfo {volume} {129}},\ \bibinfo {pages} {064111} (\bibinfo
  {year} {2008})}\BibitemShut {NoStop}%
\bibitem [{\citenamefont {Manninen}\ and\ \citenamefont {Vaara}(2006)}]{man06}%
  \BibitemOpen
  \bibfield  {author} {\bibinfo {author} {\bibfnamefont {M.}~\bibnamefont
  {Manninen}}\ and\ \bibinfo {author} {\bibfnamefont {J.}~\bibnamefont
  {Vaara}},\ }\href@noop {} {\bibfield  {journal} {\bibinfo  {journal} {J.
  Comput. Chem.}\ }\textbf {\bibinfo {volume} {27}},\ \bibinfo {pages} {434}
  (\bibinfo {year} {2006})}\BibitemShut {NoStop}%
\bibitem [{\citenamefont {Roukala}\ \emph {et~al.}(2015)\citenamefont
  {Roukala}, \citenamefont {Zhu}, \citenamefont {Giri}, \citenamefont
  {Rissanen}, \citenamefont {Lantto},\ and\ \citenamefont
  {Telkki}}]{roukala_2015}%
  \BibitemOpen
  \bibfield  {author} {\bibinfo {author} {\bibfnamefont {J.}~\bibnamefont
  {Roukala}}, \bibinfo {author} {\bibfnamefont {J.}~\bibnamefont {Zhu}},
  \bibinfo {author} {\bibfnamefont {C.}~\bibnamefont {Giri}}, \bibinfo {author}
  {\bibfnamefont {K.}~\bibnamefont {Rissanen}}, \bibinfo {author}
  {\bibfnamefont {P.}~\bibnamefont {Lantto}}, \ and\ \bibinfo {author}
  {\bibfnamefont {V.-V.}\ \bibnamefont {Telkki}},\ }\href@noop {} {\bibfield
  {journal} {\bibinfo  {journal} {J. Am. Chem. Soc.}\ }\textbf {\bibinfo
  {volume} {137}},\ \bibinfo {pages} {2464} (\bibinfo {year}
  {2015})}\BibitemShut {NoStop}%
\bibitem [{\citenamefont {Auer}\ and\ \citenamefont
  {Gauss}(2009{\natexlab{a}})}]{aue01}%
  \BibitemOpen
  \bibfield  {author} {\bibinfo {author} {\bibfnamefont {A.~A.}\ \bibnamefont
  {Auer}}\ and\ \bibinfo {author} {\bibfnamefont {J.}~\bibnamefont {Gauss}},\
  }\href@noop {} {\bibfield  {journal} {\bibinfo  {journal} {J. Chem. Phys.}\
  }\textbf {\bibinfo {volume} {115}},\ \bibinfo {pages} {1619} (\bibinfo {year}
  {2009}{\natexlab{a}})}\BibitemShut {NoStop}%
\bibitem [{\citenamefont {Auer}\ and\ \citenamefont
  {Gauss}(2009{\natexlab{b}})}]{aue09}%
  \BibitemOpen
  \bibfield  {author} {\bibinfo {author} {\bibfnamefont {A.~A.}\ \bibnamefont
  {Auer}}\ and\ \bibinfo {author} {\bibfnamefont {J.}~\bibnamefont {Gauss}},\
  }\href@noop {} {\bibfield  {journal} {\bibinfo  {journal} {Chem. Phys.}\
  }\textbf {\bibinfo {volume} {356}},\ \bibinfo {pages} {7} (\bibinfo {year}
  {2009}{\natexlab{b}})}\BibitemShut {NoStop}%
\bibitem [{\citenamefont {Enevoldsen}, \citenamefont {Oddershede},\ and\
  \citenamefont {Sauer}(1998)}]{ene98}%
  \BibitemOpen
  \bibfield  {author} {\bibinfo {author} {\bibfnamefont {T.}~\bibnamefont
  {Enevoldsen}}, \bibinfo {author} {\bibfnamefont {J.}~\bibnamefont
  {Oddershede}}, \ and\ \bibinfo {author} {\bibfnamefont {S.~P.~A.}\
  \bibnamefont {Sauer}},\ }\href@noop {} {\bibfield  {journal} {\bibinfo
  {journal} {Theor. Chem. Acc.}\ }\textbf {\bibinfo {volume} {100}},\ \bibinfo
  {pages} {275} (\bibinfo {year} {1998})}\BibitemShut {NoStop}%
\bibitem [{dal()}]{dalton}%
  \BibitemOpen
  \href@noop {} {}\bibinfo {note} {Dalton, a molecular electronic structure
  program, Release v2016.1 (2016), see {\tt
  http://daltonprogram.org}.}\BibitemShut {Stop}%
\bibitem [{\citenamefont {Aidas}\ \emph {et~al.}(2014)\citenamefont {Aidas},
  \citenamefont {Angeli}, \citenamefont {Bak}, \citenamefont {Bakken},
  \citenamefont {Bast}, \citenamefont {Boman}, \citenamefont {Christiansen},
  \citenamefont {Cimiraglia}, \citenamefont {Coriani}, \citenamefont {Dahle},
  \citenamefont {Dalskov}, \citenamefont {Ekstr\"om}, \citenamefont
  {Enevoldsen}, \citenamefont {Eriksen}, \citenamefont {Ettenhuber},
  \citenamefont {Fern\'andez}, \citenamefont {Ferrighi}, \citenamefont
  {Fliegl}, \citenamefont {Frediani}, \citenamefont {Hald}, \citenamefont
  {Halkier}, \citenamefont {H\"attig}, \citenamefont {Heiberg}, \citenamefont
  {Helgaker}, \citenamefont {Hennum}, \citenamefont {Hettema}, \citenamefont
  {Hjerten{\ae}s}, \citenamefont {H{\oe}st}, \citenamefont {H{\oe}yvik},
  \citenamefont {Iozzi}, \citenamefont {Jans{\'i}k}, \citenamefont {Jensen},
  \citenamefont {Jonsson}, \citenamefont {J{\oe}rgensen}, \citenamefont
  {Kauczor}, \citenamefont {Kirpekar}, \citenamefont {Kj{\ae}rgaard},
  \citenamefont {Klopper}, \citenamefont {Knecht}, \citenamefont {Kobayashi},
  \citenamefont {Koch}, \citenamefont {Kongsted}, \citenamefont {Krapp},
  \citenamefont {Kristensen}, \citenamefont {Ligabue}, \citenamefont
  {Lutn{\ae}s}, \citenamefont {Melo}, \citenamefont {Mikkelsen}, \citenamefont
  {Myhre}, \citenamefont {Neiss}, \citenamefont {Nielsen}, \citenamefont
  {Norman}, \citenamefont {Olsen}, \citenamefont {Olsen}, \citenamefont
  {Osted}, \citenamefont {Packer}, \citenamefont {Pawlowski}, \citenamefont
  {Pedersen}, \citenamefont {Provasi}, \citenamefont {Reine}, \citenamefont
  {Rinkevicius}, \citenamefont {Ruden}, \citenamefont {Ruud}, \citenamefont
  {Rybkin}, \citenamefont {Sa{\l}ek}, \citenamefont {Samson}, \citenamefont
  {S\'anchez~de Mer\'as}, \citenamefont {Saue}, \citenamefont {Sauer},
  \citenamefont {Schimmelpfennig}, \citenamefont {Sneskov}, \citenamefont
  {Steindal}, \citenamefont {Sylvester-Hvid}, \citenamefont {Taylor},
  \citenamefont {Teale}, \citenamefont {Tellgren}, \citenamefont {Tew},
  \citenamefont {Thorvaldsen}, \citenamefont {Th{\oe}gersen}, \citenamefont
  {Vahtras}, \citenamefont {Watson}, \citenamefont {Wilson}, \citenamefont
  {Ziolkowski},\ and\ \citenamefont {{\AA}gren}}]{aid14}%
  \BibitemOpen
  \bibfield  {author} {\bibinfo {author} {\bibfnamefont {K.}~\bibnamefont
  {Aidas}}, \bibinfo {author} {\bibfnamefont {C.}~\bibnamefont {Angeli}},
  \bibinfo {author} {\bibfnamefont {K.~L.}\ \bibnamefont {Bak}}, \bibinfo
  {author} {\bibfnamefont {V.}~\bibnamefont {Bakken}}, \bibinfo {author}
  {\bibfnamefont {R.}~\bibnamefont {Bast}}, \bibinfo {author} {\bibfnamefont
  {L.}~\bibnamefont {Boman}}, \bibinfo {author} {\bibfnamefont
  {O.}~\bibnamefont {Christiansen}}, \bibinfo {author} {\bibfnamefont
  {R.}~\bibnamefont {Cimiraglia}}, \bibinfo {author} {\bibfnamefont
  {S.}~\bibnamefont {Coriani}}, \bibinfo {author} {\bibfnamefont
  {P.}~\bibnamefont {Dahle}}, \bibinfo {author} {\bibfnamefont {E.~K.}\
  \bibnamefont {Dalskov}}, \bibinfo {author} {\bibfnamefont {U.}~\bibnamefont
  {Ekstr\"om}}, \bibinfo {author} {\bibfnamefont {T.}~\bibnamefont
  {Enevoldsen}}, \bibinfo {author} {\bibfnamefont {J.~J.}\ \bibnamefont
  {Eriksen}}, \bibinfo {author} {\bibfnamefont {P.}~\bibnamefont {Ettenhuber}},
  \bibinfo {author} {\bibfnamefont {B.}~\bibnamefont {Fern\'andez}}, \bibinfo
  {author} {\bibfnamefont {L.}~\bibnamefont {Ferrighi}}, \bibinfo {author}
  {\bibfnamefont {H.}~\bibnamefont {Fliegl}}, \bibinfo {author} {\bibfnamefont
  {L.}~\bibnamefont {Frediani}}, \bibinfo {author} {\bibfnamefont
  {K.}~\bibnamefont {Hald}}, \bibinfo {author} {\bibfnamefont {A.}~\bibnamefont
  {Halkier}}, \bibinfo {author} {\bibfnamefont {C.}~\bibnamefont {H\"attig}},
  \bibinfo {author} {\bibfnamefont {H.}~\bibnamefont {Heiberg}}, \bibinfo
  {author} {\bibfnamefont {T.}~\bibnamefont {Helgaker}}, \bibinfo {author}
  {\bibfnamefont {A.~C.}\ \bibnamefont {Hennum}}, \bibinfo {author}
  {\bibfnamefont {H.}~\bibnamefont {Hettema}}, \bibinfo {author} {\bibfnamefont
  {E.}~\bibnamefont {Hjerten{\ae}s}}, \bibinfo {author} {\bibfnamefont
  {S.}~\bibnamefont {H{\oe}st}}, \bibinfo {author} {\bibfnamefont {I.-M.}\
  \bibnamefont {H{\oe}yvik}}, \bibinfo {author} {\bibfnamefont {M.~F.}\
  \bibnamefont {Iozzi}}, \bibinfo {author} {\bibfnamefont {B.}~\bibnamefont
  {Jans{\'i}k}}, \bibinfo {author} {\bibfnamefont {H.~J.~{\relax Aa}.}\
  \bibnamefont {Jensen}}, \bibinfo {author} {\bibfnamefont {D.}~\bibnamefont
  {Jonsson}}, \bibinfo {author} {\bibfnamefont {P.}~\bibnamefont
  {J{\oe}rgensen}}, \bibinfo {author} {\bibfnamefont {J.}~\bibnamefont
  {Kauczor}}, \bibinfo {author} {\bibfnamefont {S.}~\bibnamefont {Kirpekar}},
  \bibinfo {author} {\bibfnamefont {T.}~\bibnamefont {Kj{\ae}rgaard}}, \bibinfo
  {author} {\bibfnamefont {W.}~\bibnamefont {Klopper}}, \bibinfo {author}
  {\bibfnamefont {S.}~\bibnamefont {Knecht}}, \bibinfo {author} {\bibfnamefont
  {R.}~\bibnamefont {Kobayashi}}, \bibinfo {author} {\bibfnamefont
  {H.}~\bibnamefont {Koch}}, \bibinfo {author} {\bibfnamefont {J.}~\bibnamefont
  {Kongsted}}, \bibinfo {author} {\bibfnamefont {A.}~\bibnamefont {Krapp}},
  \bibinfo {author} {\bibfnamefont {K.}~\bibnamefont {Kristensen}}, \bibinfo
  {author} {\bibfnamefont {A.}~\bibnamefont {Ligabue}}, \bibinfo {author}
  {\bibfnamefont {O.~B.}\ \bibnamefont {Lutn{\ae}s}}, \bibinfo {author}
  {\bibfnamefont {J.~I.}\ \bibnamefont {Melo}}, \bibinfo {author}
  {\bibfnamefont {K.~V.}\ \bibnamefont {Mikkelsen}}, \bibinfo {author}
  {\bibfnamefont {R.~H.}\ \bibnamefont {Myhre}}, \bibinfo {author}
  {\bibfnamefont {C.}~\bibnamefont {Neiss}}, \bibinfo {author} {\bibfnamefont
  {C.~B.}\ \bibnamefont {Nielsen}}, \bibinfo {author} {\bibfnamefont
  {P.}~\bibnamefont {Norman}}, \bibinfo {author} {\bibfnamefont
  {J.}~\bibnamefont {Olsen}}, \bibinfo {author} {\bibfnamefont {J.~M.~H.}\
  \bibnamefont {Olsen}}, \bibinfo {author} {\bibfnamefont {A.}~\bibnamefont
  {Osted}}, \bibinfo {author} {\bibfnamefont {M.~J.}\ \bibnamefont {Packer}},
  \bibinfo {author} {\bibfnamefont {F.}~\bibnamefont {Pawlowski}}, \bibinfo
  {author} {\bibfnamefont {T.~B.}\ \bibnamefont {Pedersen}}, \bibinfo {author}
  {\bibfnamefont {P.~F.}\ \bibnamefont {Provasi}}, \bibinfo {author}
  {\bibfnamefont {S.}~\bibnamefont {Reine}}, \bibinfo {author} {\bibfnamefont
  {Z.}~\bibnamefont {Rinkevicius}}, \bibinfo {author} {\bibfnamefont {T.~A.}\
  \bibnamefont {Ruden}}, \bibinfo {author} {\bibfnamefont {K.}~\bibnamefont
  {Ruud}}, \bibinfo {author} {\bibfnamefont {V.~V.}\ \bibnamefont {Rybkin}},
  \bibinfo {author} {\bibfnamefont {P.}~\bibnamefont {Sa{\l}ek}}, \bibinfo
  {author} {\bibfnamefont {C.~C.~M.}\ \bibnamefont {Samson}}, \bibinfo {author}
  {\bibfnamefont {A.}~\bibnamefont {S\'anchez~de Mer\'as}}, \bibinfo {author}
  {\bibfnamefont {T.}~\bibnamefont {Saue}}, \bibinfo {author} {\bibfnamefont
  {S.~P.~A.}\ \bibnamefont {Sauer}}, \bibinfo {author} {\bibfnamefont
  {B.}~\bibnamefont {Schimmelpfennig}}, \bibinfo {author} {\bibfnamefont
  {K.}~\bibnamefont {Sneskov}}, \bibinfo {author} {\bibfnamefont {A.~H.}\
  \bibnamefont {Steindal}}, \bibinfo {author} {\bibfnamefont {K.~O.}\
  \bibnamefont {Sylvester-Hvid}}, \bibinfo {author} {\bibfnamefont {P.~R.}\
  \bibnamefont {Taylor}}, \bibinfo {author} {\bibfnamefont {A.~M.}\
  \bibnamefont {Teale}}, \bibinfo {author} {\bibfnamefont {E.~I.}\ \bibnamefont
  {Tellgren}}, \bibinfo {author} {\bibfnamefont {D.~P.}\ \bibnamefont {Tew}},
  \bibinfo {author} {\bibfnamefont {A.~J.}\ \bibnamefont {Thorvaldsen}},
  \bibinfo {author} {\bibfnamefont {L.}~\bibnamefont {Th{\oe}gersen}}, \bibinfo
  {author} {\bibfnamefont {O.}~\bibnamefont {Vahtras}}, \bibinfo {author}
  {\bibfnamefont {M.~A.}\ \bibnamefont {Watson}}, \bibinfo {author}
  {\bibfnamefont {D.~J.~D.}\ \bibnamefont {Wilson}}, \bibinfo {author}
  {\bibfnamefont {M.}~\bibnamefont {Ziolkowski}}, \ and\ \bibinfo {author}
  {\bibfnamefont {H.}~\bibnamefont {{\AA}gren}},\ }\href@noop {} {\bibfield
  {journal} {\bibinfo  {journal} {WIREs Comput. Mol. Sci.}\ }\textbf {\bibinfo
  {volume} {4}},\ \bibinfo {pages} {269} (\bibinfo {year} {2014})}\BibitemShut
  {NoStop}%
\bibitem [{\citenamefont {Adamo}\ and\ \citenamefont
  {Barone}(1998)}]{adamo_1998}%
  \BibitemOpen
  \bibfield  {author} {\bibinfo {author} {\bibfnamefont {C.}~\bibnamefont
  {Adamo}}\ and\ \bibinfo {author} {\bibfnamefont {V.}~\bibnamefont {Barone}},\
  }\href@noop {} {\bibfield  {journal} {\bibinfo  {journal} {Chem. Phys.
  Lett.}\ }\textbf {\bibinfo {volume} {298}},\ \bibinfo {pages} {113} (\bibinfo
  {year} {1998})}\BibitemShut {NoStop}%
\bibitem [{\citenamefont {Adamo}\ and\ \citenamefont
  {Barone}(1999)}]{adamo_1999}%
  \BibitemOpen
  \bibfield  {author} {\bibinfo {author} {\bibfnamefont {C.}~\bibnamefont
  {Adamo}}\ and\ \bibinfo {author} {\bibfnamefont {V.}~\bibnamefont {Barone}},\
  }\href@noop {} {\bibfield  {journal} {\bibinfo  {journal} {J. Chem. Phys.}\
  }\textbf {\bibinfo {volume} {110}},\ \bibinfo {pages} {6158} (\bibinfo {year}
  {1999})}\BibitemShut {NoStop}%
\bibitem [{per()}]{perdew_1996}%
  \BibitemOpen
  \href@noop {} {}\bibinfo {note} {{J.~P.~Perdew, K.~Burke, and M.~Ernzerhof,
  Phys.\ Rev.\ Lett.\ {\bf 77}, 3865 (1996); {\em Ibid.}\ {\bf 78}, 1396(E)
  (1997)}}\BibitemShut {NoStop}%
\bibitem [{\citenamefont {Becke}(1988)}]{becke_1988}%
  \BibitemOpen
  \bibfield  {author} {\bibinfo {author} {\bibfnamefont {A.~D.}\ \bibnamefont
  {Becke}},\ }\href@noop {} {\bibfield  {journal} {\bibinfo  {journal} {Phys.
  Rev. A}\ }\textbf {\bibinfo {volume} {38}},\ \bibinfo {pages} {3098}
  (\bibinfo {year} {1988})}\BibitemShut {NoStop}%
\bibitem [{\citenamefont {Lee}, \citenamefont {Yang},\ and\ \citenamefont
  {Parr}(1988)}]{lee_1988}%
  \BibitemOpen
  \bibfield  {author} {\bibinfo {author} {\bibfnamefont {C.}~\bibnamefont
  {Lee}}, \bibinfo {author} {\bibfnamefont {W.}~\bibnamefont {Yang}}, \ and\
  \bibinfo {author} {\bibfnamefont {R.~G.}\ \bibnamefont {Parr}},\ }\href@noop
  {} {\bibfield  {journal} {\bibinfo  {journal} {Phys. Rev. B}\ }\textbf
  {\bibinfo {volume} {37}},\ \bibinfo {pages} {785} (\bibinfo {year}
  {1988})}\BibitemShut {NoStop}%
\bibitem [{\citenamefont {Miehlich}\ \emph {et~al.}(1989)\citenamefont
  {Miehlich}, \citenamefont {Savin}, \citenamefont {Stoll},\ and\ \citenamefont
  {Preuss}}]{miehlich_1989}%
  \BibitemOpen
  \bibfield  {author} {\bibinfo {author} {\bibfnamefont {B.}~\bibnamefont
  {Miehlich}}, \bibinfo {author} {\bibfnamefont {A.}~\bibnamefont {Savin}},
  \bibinfo {author} {\bibfnamefont {H.}~\bibnamefont {Stoll}}, \ and\ \bibinfo
  {author} {\bibfnamefont {H.}~\bibnamefont {Preuss}},\ }\href@noop {}
  {\bibfield  {journal} {\bibinfo  {journal} {Chem. Phys. Lett.}\ }\textbf
  {\bibinfo {volume} {157}},\ \bibinfo {pages} {200} (\bibinfo {year}
  {1989})}\BibitemShut {NoStop}%
\bibitem [{\citenamefont {Vosko}, \citenamefont {Wilk},\ and\ \citenamefont
  {Nusair}(1980)}]{vosko_1980}%
  \BibitemOpen
  \bibfield  {author} {\bibinfo {author} {\bibfnamefont {S.~H.}\ \bibnamefont
  {Vosko}}, \bibinfo {author} {\bibfnamefont {L.}~\bibnamefont {Wilk}}, \ and\
  \bibinfo {author} {\bibfnamefont {M.}~\bibnamefont {Nusair}},\ }\href@noop {}
  {\bibfield  {journal} {\bibinfo  {journal} {Can. J. Phys.}\ }\textbf
  {\bibinfo {volume} {58}},\ \bibinfo {pages} {1200} (\bibinfo {year}
  {1980})}\BibitemShut {NoStop}%
\bibitem [{\citenamefont {Becke}(1993)}]{becke_1993}%
  \BibitemOpen
  \bibfield  {author} {\bibinfo {author} {\bibfnamefont {A.~D.}\ \bibnamefont
  {Becke}},\ }\href@noop {} {\bibfield  {journal} {\bibinfo  {journal} {J.
  Chem. Phys.}\ }\textbf {\bibinfo {volume} {98}},\ \bibinfo {pages} {5648}
  (\bibinfo {year} {1993})}\BibitemShut {NoStop}%
\bibitem [{\citenamefont {Stephens}\ \emph {et~al.}(1994)\citenamefont
  {Stephens}, \citenamefont {Devlin}, \citenamefont {Chabalowski},\ and\
  \citenamefont {Frisch}}]{stephens_1994}%
  \BibitemOpen
  \bibfield  {author} {\bibinfo {author} {\bibfnamefont {P.~J.}\ \bibnamefont
  {Stephens}}, \bibinfo {author} {\bibfnamefont {F.~J.}\ \bibnamefont
  {Devlin}}, \bibinfo {author} {\bibfnamefont {C.~F.}\ \bibnamefont
  {Chabalowski}}, \ and\ \bibinfo {author} {\bibfnamefont {M.~J.}\ \bibnamefont
  {Frisch}},\ }\href@noop {} {\bibfield  {journal} {\bibinfo  {journal} {J.
  Phys. Chem.}\ }\textbf {\bibinfo {volume} {98}},\ \bibinfo {pages} {11623}
  (\bibinfo {year} {1994})}\BibitemShut {NoStop}%
\bibitem [{\citenamefont {Sun}, \citenamefont {Liu},\ and\ \citenamefont
  {Kutzelnigg}(2011)}]{sun11}%
  \BibitemOpen
  \bibfield  {author} {\bibinfo {author} {\bibfnamefont {Q.}~\bibnamefont
  {Sun}}, \bibinfo {author} {\bibfnamefont {W.}~\bibnamefont {Liu}}, \ and\
  \bibinfo {author} {\bibfnamefont {W.}~\bibnamefont {Kutzelnigg}},\
  }\href@noop {} {\bibfield  {journal} {\bibinfo  {journal} {Theor. Chem.
  Acc.}\ }\textbf {\bibinfo {volume} {129}},\ \bibinfo {pages} {423} (\bibinfo
  {year} {2011})}\BibitemShut {NoStop}%
\bibitem [{\citenamefont {Terrano}()}]{ter19}%
  \BibitemOpen
  \bibfield  {author} {\bibinfo {author} {\bibfnamefont {W.~A.}\ \bibnamefont
  {Terrano}},\ }\href@noop {} {}\bibinfo {note} {{\rm private communication,
  2018.}}\BibitemShut {Stop}%
\bibitem [{\citenamefont {Hanni}\ \emph {et~al.}(2004)\citenamefont {Hanni},
  \citenamefont {Lantto}, \citenamefont {Runeberg}, \citenamefont {Jokisaari},\
  and\ \citenamefont {Vaara}}]{hanni_2004}%
  \BibitemOpen
  \bibfield  {author} {\bibinfo {author} {\bibfnamefont {M.}~\bibnamefont
  {Hanni}}, \bibinfo {author} {\bibfnamefont {P.}~\bibnamefont {Lantto}},
  \bibinfo {author} {\bibfnamefont {N.}~\bibnamefont {Runeberg}}, \bibinfo
  {author} {\bibfnamefont {J.}~\bibnamefont {Jokisaari}}, \ and\ \bibinfo
  {author} {\bibfnamefont {J.}~\bibnamefont {Vaara}},\ }\href@noop {}
  {\bibfield  {journal} {\bibinfo  {journal} {J. Chem. Phys.}\ }\textbf
  {\bibinfo {volume} {121}},\ \bibinfo {pages} {5908} (\bibinfo {year}
  {2004})}\BibitemShut {NoStop}%
\end{thebibliography}%
\bibliographystyle{aipnum4-1}

%\newpage\noindent
%{\Large\bf Graphical TOC Entry}\\
%\begin{tocentry}
%\begin{center}
%\includegraphics[width=8.25cm]{toc.jpg}
%\end{center}
%\end{tocentry}

\end{document}